\documentclass[twocolumn,twocolappendix,tighten,times,astrosymb]{aastex7}

\usepackage{mathtools}
\renewcommand{\vec}[1]{\mathbf{#1}}
\usepackage[T1]{fontenc}
\usepackage{microtype}

\shorttitle{\small Gro\v selj et al.}
\shortauthors{\small Gro\v selj et al.}
\begin{document}

\title{\large High-energy Emission from 
Turbulent Electron-ion Coronae of Accreting Black Holes}

\author[orcid=0000-0002-5408-3046]{Daniel Gro\v selj}
\affiliation{Centre for mathematical Plasma Astrophysics, Department of Mathematics, 
KU Leuven, B-3001 Leuven, Belgium}
\email[show]{daniel.grosel@gmail.com}
\correspondingauthor{Daniel Gro\v selj}
\author[orcid=0000-0001-7801-0362]{Alexander Philippov}
\affiliation{Department of Physics, University of Maryland, College Park, MD 20742, USA}
\email{sashaph@umd.edu}
\author[orcid=0000-0001-5660-3175]{Andrei M.~Beloborodov}
\affiliation{Physics Department and Columbia Astrophysics Laboratory, Columbia University, New York, NY 10027, USA}
\affiliation{Max Planck Institute for Astrophysics, D-85741 Garching, Germany}
\email{amb2046@columbia.edu}
\author[orcid=0000-0002-7962-5446]{Richard Mushotzky}
\affiliation{Department of Astronomy and Joint Space-Science Institute, University of Maryland, College Park, MD 20742, USA}
\email{rmushotz@umd.edu}

\begin{abstract}
We develop a model of particle energization and emission from strongly turbulent black-hole coronae. 
Our local model is based on a set of 2D radiative particle-in-cell simulations with an electron-ion plasma composition, injection 
and diffusive escape of photons and charged particles, and self-consistent Compton scattering. 
We show that a radiatively compact turbulent corona generates extended nonthermal ion distributions, 
while producing X-ray spectra consistent with 
observations. As an example, we demonstrate excellent agreement with 
observed X-ray spectra of NGC 4151. The predicted 
emission spectra feature an MeV tail, which can be studied with future MeV-band instruments. The MeV tail 
is shaped by nonthermal electrons accelerated at turbulent current sheets.
We also find that the corona regulates itself into a two-temperature state, with ions
much hotter than electrons. The ions carry away roughly two-thirds of the dissipated power, and their energization
is driven by a combination of shocks and reconnecting current sheets,
embedded into the turbulent flow.
\end{abstract}


\section{Introduction}
Supermassive accreting black holes in active galactic nuclei (AGNs) 
are among the most powerful 
sources of hard X-rays \citep{Laha2024,Kara2025}.
Based on their prevalence and physical properties, 
they are believed to be also among the prime 
sources of high-energy cosmic 
rays and neutrinos \citep{Inoue2019,Murase2020,Padovani2024,IceCube2024}. 
The hard X-ray emission from AGNs is commonly attributed to a spatially compact region
near the black hole, referred to as the ``corona.'' The dimensionless 
ratio of the coronal luminosity $L$ to its size $R$ defines the radiative 
compactness $\ell = 4\pi (m_{\rm p}/m_{\rm e}) (L/L_{\rm E}) (R / R_{\rm g})^{-1}$
\citep{Guilbert1983}, where $L_{\rm E}$ is the 
Eddington limit, $R_{\rm g}$ is the gravitational radius, 
and $m_{\rm p}/m_{\rm e}$ is the proton-electron mass ratio.
Many observed accreting black holes have $\ell\gg 1$,
which implies that black-hole coronae are radiatively dense environments 
where matter and radiation are 
strongly coupled. Most notably, the high-compactness regime is characterized by 
moderate optical depths, fast radiative cooling, and (in the 
most extreme cases) electron-positron 
pair creation \citep[e.g.,][]{Svensson1984,Fabian2015,Beloborodov2017}.

A well-established view is that the coronal X-ray emission stems from the Compton 
scattering of low-energy seed photons off hot electrons
with an effective temperature of about $100$ keV
\citep{Shapiro1976, Haardt1993}. The 
rapid cooling experienced by the Comptonizing electrons points to
an efficient form of coronal energy dissipation, which is needed to 
maintain the electrons at the observationally inferred temperatures.

Much insight into the physics of black-hole coronae 
can be gained by considering \emph{how} the electrons are energized. 
The physics of electron energization involves collective plasma interactions, 
such as turbulence and/or magnetic reconnection \citep{Petrosian2012,Lazarian2012,Guo2024,Sironi2025}, 
which release the energy stored in the coronal magnetic fields
and bulk motions into heat, nonthermal particles, and radiation. Thus, 
studies of plasma dissipation under the conditions that reproduce the 
observed coronal X-ray signal 
can place important constraints on the physics 
of the emission mechanism.

The same process which
powers the electron energization, and in turn the X-ray emission, determines also
the energetics of the coronal ions (i.e., protons). 
The composition of black-hole coronae is presently not well known; it could be either leptonic 
and dominated by pairs or of the hadronic type with a substantial amount of protons.
In a number of sources, the compactness $\ell$ may not be high
enough to sustain a pair-plasma-dominated state \citep{Haardt1997,Hinkle2021}. Models focusing on electron-ion
compositions are particularly timely, given the recent neutrino observations by IceCube \citep{IceCube2024}, which 
suggest that AGN coronae may be efficient proton accelerators.
How the released energy is partitioned between the escaping photons and cosmic-ray protons 
can be investigated in particle-in-cell (PIC) simulations with explicit radiative transfer, which track the 
self-consistent interactions among photons, charged particles and their 
electromagnetic fields \citep{Groselj2024, Nattila2024}. However, existing studies focus mainly
on pair plasma compositions.

In this paper, we study the partitioning of energy among photons and charged particles in 
strongly turbulent \emph{electron-ion} coronae of accreting black holes. 
To this end, we perform the first PIC simulations with self-consistent Compton scattering
in an electron-ion turbulent plasma. Alongside the numerical results we develop theoretical estimates.
The predicted X-ray spectra are compared with observations of NGC 4151---one of the
X-ray brightest AGNs \citep{Ulrich2000} and 
a candidate neutrino source \citep{Inoue2023, Neronov2024, IceCube2024, Murase2024}.
As we demonstrate below, the regime of electron-ion 
turbulence applicable to black-hole coronae features trans-sonic and trans-Alfv\' enic
motions, leading to sporadic formation of magnetized shocks. 
The shocks, alongside reconnecting current sheets, act 
as drivers of particle heating and acceleration. The electron
energization is largely offset by the rapid radiative cooling, which results in 
the formation of a two-temperature corona with ions much 
hotter than electrons ($T_{\rm i} \gg T_{\rm e}$). When the corona
is sufficiently magnetized, we find that ions form extended nonthermal energy spectra. 
Our results have direct implications
for the production of X-ray photons, cosmic rays and neutrinos in coronae
of supermassive black holes.

\section{Theoretical Expectations}
\label{sec:theory}

We consider a scenario where the coronal dissipation is mediated by magnetized turbulence in a weakly 
collisional electron-proton plasma.\footnote{Our analytical estimates focus on the case where the 
plasma ions are protons with mass $m_{\rm i}\approx 1836\,m_{\rm e}$. However, the simulations (Sec.~\ref{sec:simulations}) 
employ an artificially reduced ion mass of $m_{\rm i}=144\,m_{\rm e}$ for computational convenience.}
Turbulently energized charged particles and photons escape the source via spatial diffusion, 
while at the same time low-energy thermal plasma and seed photons are resupplied into the domain. Energy released into the 
electrons is quickly passed onto photons and radiated away, while the fraction of energy received by the protons 
goes into heating and nonthermal acceleration. Below, we provide analytical estimates in support of this picture.

\subsection{Regime of Turbulence}
\label{sec:turb_theory}
The global shape and multiphase structure of black-hole coronae, 
including the field line geometry, is presently the subject of 
ongoing investigations \citep[e.g.,][]{Jiang2019, Chaskina2021, Liska2022, Liska2023, Scepi2024, 
Sridhar2025, Krawczynski2025, Nagele2026}. On the other hand, from the point of view of a local model, important constraints on 
the regime of turbulence can be obtained based on two aspects: i) the observationally inferred large radiative 
compactness $\ell \gg 1$, and ii) the presumed efficient particle acceleration. Alongside recent 
IceCube observations \citep{IceCube2024}, hinting at proton acceleration, measurements of an extended MeV emission tail in 
the stellar-mass binary Cygnus X-1 \citep{McConnell2002,Zdziarski2017} imply that black-hole coronae can
also accelerate electrons into nonthermal power laws.

\paragraph{Amplitude of turbulent fluctuations}Recent studies of kinetic turbulence in local boxes with a mean magnetic field $B_0$ showed that 
efficient nonthermal particle acceleration strongly favors the large-amplitude turbulence regime with
typical magnetic fluctuation strength $\delta B/B_0 \gtrsim 1$ \citep{Comisso2018,Nattila2022,Vega2024}. 
Since we aim to uncover regimes of efficient particle acceleration, we focus here on the case $\delta B/B_0\sim 1$.

\paragraph{Radiative compactness}For our model, the compactness $\ell$ can be expressed as follows. 
In steady state, the power channeled into escaping radiation per unit volume of a 
turbulent electron-ion plasma is $\dot{U}_{\rm esc} \simeq (1 - q_{\rm i})B_0^2 v_{\rm A}/ (8\pi s_0)$,
where $q_{\rm i}$ is the \emph{ion heating fraction}, $s_0$ is the turbulence driving scale, and $v_{\rm A}$ is the Alfv\' en speed. 
The fraction $q_{\rm i}$ represents the relative amount of dissipated power 
transferred to the (non-radiating) ions.\footnote{For historic reasons, we refer to $q_{\rm i}$ colloquially as the ``heating'' 
fraction, even though $q_{\rm i}$ is defined based on the total power channeled into ions, 
be it in the form of thermal or nonthermal particles.} 
In a cubic slab with linear size $S$, the compactness can be expressed as 
$\ell = S^2\dot{U}_{\rm esc} \sigma_{\rm T} / (m_{\rm e}c^3)$ \citep[e.g.,][]{Stern1995}. This gives
\begin{align}
    \ell \simeq 2\tau_{\rm T}\sigma_{\rm e}(1 - q_{\rm i})(v_{\rm A}/c),
    \label{eq:lrad1}
\end{align}
where $\tau_{\rm T} = \sigma_{\rm T} n_{\rm 0} s_{\rm esc}$ is the Thomson optical
depth, $n_{0}$ is the mean electron (or ion) density, 
$\sigma_{\rm e} = B_0^2 / (4\pi n_{0} m_{\rm e} c^2)$ is the electron 
magnetization, and $s_{\rm esc} \simeq S/2$ is the typical distance over which photons need to 
diffuse in order to escape the source.
To obtain the above estimate, we 
assumed that turbulence is excited on the global scale, 
such that $s_0 \simeq s_{\rm esc}$. For simplicity, we will retain this assumption throughout 
the rest of the paper, but we note that the ratio $s_{\rm esc}/s_0$ can be in principle 
kept as a free parameter of the model.\footnote{Our focus on the regime with $s_{\rm esc}\simeq s_0$ is also dictated 
by computational constraints. In particular, at larger values of $s_{\rm esc}/s_0$ the simulations become increasingly 
more challenging because it takes more time to reach a steady state. In principle, the 
regime with $s_{\rm esc}/s_0 \gg 1$ could be quite favorable for proton acceleration (see discussion in Sec.~\ref{sec:ion_heating}).}
The ion heating fraction $q_{\rm i}$ in \eqref{eq:lrad1} is not known \emph{a priori}; it has been extensively studied 
in (reduced) kinetic simulations \citep[e.g.,][]{Kawazura2019,Kawazura2020,Zhdankin2019,Zhdankin2021,Zhdankin2021b}, but so far 
not in the regime directly applicable to strongly turbulent black-hole coronae. We argue below that $q_{\rm i}$ should be of order unity
under conditions expected in black-hole coronae, and we measure $q_{\rm i}\approx 0.6$ -- $0.7$ in 
simulations (Sec.~\ref{sec:simulations}).

\paragraph{Plasma magnetization}Equation \eqref{eq:lrad1} essentially constrains the relevant range of plasma magnetizations. 
Assuming $q_{\rm i} = 0.65$, $\tau_{\rm T} = 1$, and $m_{\rm i}/m_{\rm e} = 1836$ we find that the
observationally relevant range $1 \lesssim \ell \lesssim 100$ corresponds to:
\begin{align}
0.01 \lesssim \sigma_{\rm i} \lesssim 0.2,
\label{eq:sigma_range}
\end{align}
where $\sigma_{\rm i} = B_0^2 / (4\pi n_{0} m_{\rm i} c^2)$ is the ion magnetization.
Thus, the coronal ions are moderately magnetized whereas 
the electrons are strongly magnetized, since $\sigma_{\rm e} = \sigma_{\rm i}m_{\rm i}/m_{\rm e}\gg 1$.
Interestingly, a recent analysis of global magneto-hydrodynamic (MHD) accretion simulations found
$\sigma_{\rm i}\sim 0.1$ in regions identified as the corona \citep{Hankla2025}, which is 
compatible with our estimate \eqref{eq:sigma_range}. We also note that the collisional coupling between 
ions and electrons, which is neglected in our model, can become relevant 
on the low side of the inferred $\sigma_{\rm i}$ range (see Sec.~\ref{ref:applicability} and
Appendix~\ref{sec:collisions}).

\paragraph{A quantitative example}It may
be instructive to put the turbulence parameters into the context of an AGN corona with a given size and black-hole 
mass. The distance $s_{\rm esc}$ can be taken as 
a proxy for the typical half-width of the corona (e.g., in spherical geometry this would be the
coronal radius $R$). We can estimate
\begin{align}
    s_{\rm esc} \sim  10^{13}\!\left(\!\frac{s_{\rm esc}}{10 R_{\rm g}}\!\right)
    \!\left(\!\frac{M}{10^7\!\!\; M_\odot}\!\right)\,{\rm cm},
\end{align}
where $M$ is the black-hole mass, $M_\odot$ is the solar mass, and $R_{\rm g} = GM/c^2$ is the gravitational radius.
The turbulent cascade timescale $t_0 \sim s_{\rm esc}/v_{\rm A}$ is:
\begin{align}
    t_0 \sim \frac{s_{\rm esc}}{v_{\rm A}} \sim 10^3\left(\frac{\sigma_{\rm i}}{0.1}\right)^{\!\!\;\!\!-1/2}\!\!\left(\!\frac{s_{\rm esc}}{10 R_{\rm g}}\!\right)\!\!
    \left(\!\frac{M}{10^7\!\!\; M_\odot}\!\right)\,{\rm s}.
    \label{eq:t0}
\end{align}
In our model, $s_{\rm esc}/v_{\rm A}$ sets the typical timescale of the short-term coronal variability. Fast X-ray variability, on timescales of a few tens of kiloseconds, 
has been detected in AGNs \citep{Zhao2025}, and present observations are likely just 
upper bounds on $t_0$. For stellar-mass binary black holes the timescale \eqref{eq:t0} 
is of the  order of a millisecond, which is 
consistent with rapid flares detected in Cygnus X-1 \citep{Gierlinski2003}.
Finally, the typical magnetic field strength is
\begin{align}
    B_0 \sim 10^4\,\tau_{\rm T}^{1/2}\!\!\;\left(\frac{\sigma_{\rm i}}{0.1}\right)^{\!\!\;\!1/2}\!\left(\!\frac{s_{\rm esc}}{10 R_{\rm g}}\!\right)^{\!\!\;\!\!\!\!\;-1/2}\!\!
    \left(\!\frac{M}{10^7\!\!\; M_\odot}\!\right)^{\!\!\;\!\!\;\!\!-1/2}\!{\rm G},
\end{align}
and the corresponding coronal luminosity $L \simeq (1-q_{\rm i})B_0^2v_{\rm A}s_{\rm esc}^2/6$
(assuming a spherical geometry) is 
\begin{align}
    L \sim 10^{43}\,
    \tau_{\rm T}\!\left(\!\frac{1 \!- \! q_{\rm i}}{0.35}\!\right)\!\left(\frac{\sigma_{\rm i}}{0.1}\right)^{\!\!\;\!3/2}\!\!\left(\!\frac{s_{\rm esc}}{10 R_{\rm g}}\!\right)\!\!
    \left(\!\frac{M}{10^7\!\!\; M_\odot}\!\right)\frac{\rm erg}{s}.
\end{align}

\subsection{Ion Energization}
\label{sec:ion_heating}
Early theoretical works on the ion heating fraction 
in kinetic plasma turbulence considered a situation where the turbulent fluctuations introduce
only mild perturbations of the particle distribution on top of a thermal 
background \citep[e.g.,][]{Quataert1998,Schekochihin2009,Howes2010a}. The ion plasma beta $\beta_{\rm i} = 8\pi n_0 T_{\rm i}/B_0^2$
and the ion-electron temperature ratio $T_{\rm i}/T_{\rm e}$ can be then taken as input parameters for calculations
of the heating fraction $q_{\rm i}$ \citep[e.g.,][]{Kawazura2019,Kawazura2020,Adkins2025}.
However, the circumstances are different in open and strongly turbulent domains with $\delta B/B_0 \sim 1$. There, 
the distribution of particles entering the domain is significantly altered within a single dynamic timescale, 
and a steady state average distribution is established via a balance between the rapid 
particle energization and diffusive escape. Thus, the ion confinement time in a strongly turbulent corona 
is an important parameter for the determination of the heating partition.
In particular, we demonstrate that the confinement time is rather limited even for the moderate-energy
particles near the thermal peak of the distribution, when the turbulence is driven on a scale 
not much smaller than the system size (see Sec.~\ref{sec:acceleration}). This is further corroborated by 
a recent analysis of particle transport in high-resolution MHD simulations \citep{Kempski2025}, which shows that moderate-energy 
particles can efficiently diffuse in space by following the strongly turbulent magnetic field lines.

\paragraph{Steady-state kinetic temperature}Let us consider a situation where the ions are \emph{nominally} in temperature equilibrium  
with electrons, such that $T_{\rm i} = T_{\rm e}$. Since we focus on large-amplitude turbulence, the 
typical amplitude of turbulent motions is $\delta v \simeq v_{\rm A}$. For ions with
non-relativistic temperatures (to be justified below), the Alfv\' en speed can be defined based on the
cold ion magnetization as $v_{\rm A} = c[\sigma_{\rm i}/(1 + \sigma_{\rm i})]^{1/2}$. Note that $v_{\rm A}\simeq c\sigma_{\rm i}^{1/2}$ is a good
approximation over the relevant range (see Eq.~\eqref{eq:sigma_range}).
The typical turbulent 
Alfv\' enic Mach number is then $M_{\rm A} = \delta v / v_{\rm A}\sim 1$, and the sonic Mach number is 
$M_{\rm s} = \delta v / c_{\rm s} \gg 1$ for typical coronal temperatures of $\sim 100$ keV, 
where $c_{\rm s} = [(\Gamma_{\rm i}T_{\rm i} + \Gamma_{\rm e}T_{\rm e})/m_{\rm i}]^{1/2}$ is the sound 
speed and $\Gamma_{s}$ is the species adiabatic index.
We therefore expect formation of shocks, which reprocess and heat the plasma.\footnote{It may seem 
that $M_{\rm A}\sim 1$ is a significant limitation 
for the formation of magnetized shocks. However, when $\delta B/B_0\sim 1$ the magnetic field magnitude $|B|$ 
exhibits order unity fluctuations, and so does the local Alfv\' en speed. As shown in Sec.~\ref{sec:turbulence},
this enables the formation of regions in space where $M_{\rm A}\sim$ a few, into which the shocks can spread.} 
The electron heating 
will be offset by their rapid radiative cooling. On the other hand, the ion kinetic temperature will rise
until the system reaches a quasi-steady state. Here, we conjecture that the ions are pinned in steady state to the trans-sonic 
regime with $\delta v \sim v_{\rm A} \sim c_{\rm s}$ and $T_{\rm i} \gg T_{\rm e}$. Since cold plasma is 
continuously resupplied into the domain, shocks form sporadically in the quasi-steady state. 
In a trans-sonic and trans-Alfv\' enic flow, the average ion kinetic temperature and ion-electron temperature 
ratio are given by:
\begin{align}
    \theta_{\rm i} \simeq \frac{\sigma_{\rm i}}{\Gamma_{\rm i}}, && 
    \frac{T_{\rm i}}{T_{\rm e}} \simeq \frac{\sigma_{\rm e}}{\Gamma_{\rm i}\theta_{\rm e}},
    \label{eq:Ti}
\end{align}
where the ion adiabatic index $\Gamma_{\rm i}\approx 5/3$ and
$\theta_{s} = T_{s}/(m_{s}c^2)$ is the dimensionless species kinetic temperature. 
The average ion kinetic temperature is therefore non-relativistic or 
very mildly relativistic, which is confirmed by simulations (Sec.~\ref{sec:simulations}). 
Essentially, the trans-sonic state with $M_{\rm s}\sim 1$ is considered here as 
an attractor of the dynamics. States with $M_{\rm s} \gg 1$ are circumvented by rapid shock heating, 
while $M_{\rm s} \ll 1$ is prevented by the limited ion 
confinement time (of the order of $\sim S/v_{\rm A}$ in our model). We emphasize that in the steady state shocks need not 
be the main energization mechanism, since on average the flow is only trans-sonic. Other dissipation channels can also operate.

\paragraph{Heating fraction}
Following the above arguments, $q_{\rm i}$ corresponds to the fraction of turbulence 
power required to maintain the steady-state ion kinetic temperature given by \eqref{eq:Ti}.
To obtain $q_{\rm i}$, we assume that ions enter the corona via turbulent advection from a cold 
external medium, so that their initial energy is roughly the 
kinetic energy of the turbulent $E\times B$ drift motion; this situation closely resembles 
our numerical model (Sec.~\ref{sec:method}). The mean ion energy gain is then comparable to their
steady state internal energy. Balancing the turbulence power input with the
rate of energy gain we obtain: $(3/2)n_0T_{\rm i}/t_{\rm conf}\simeq q_{\rm i}B_0^2 v_{\rm A}/(8\pi s_0)$,
where $t_{\rm conf}$ is the average confinement time or lifetime of ions in the corona.
In large-amplitude turbulence driven on the box scale, the expected confinement time is 
of the order of $S/v_{\rm A}$ \citep{Gorbunov2025}. More specifically, in our 
simulations with $s_0 \simeq s_{\rm esc} = S/2$
we measure $t_{\rm conf}\simeq 2 S/v_{\rm A}$ for a range 
of $\sigma_{\rm i}$ (see Sec.~\ref{sec:acceleration}). Using this result
together with \eqref{eq:Ti} we can estimate $q_{\rm i}$ from the energy balance equation as
\begin{align}
    q_{\rm i} \simeq \frac{3}{4\Gamma_{\rm i}} \simeq 0.5.
    \label{eq:qi}
\end{align}
Therefore, we expect that the escaping ions carry away 
an order-unity fraction of the dissipated power. This is confirmed by our simulations presented 
in Sec.~\ref{sec:simulations}. 

\paragraph{Nonthermal acceleration}Above the thermal peak of the distribution (see Eq.~\eqref{eq:Ti}),
the ion energy spectrum may exhibit a nonthermal tail of the form ${\rm d}N/{\rm d} E \propto E^{-p}$, where
$p$ is the power-law index. In the energy range above the thermal peak but below the TeV band, 
a hard nonthermal proton slope (with $1\lesssim p\lesssim 2$) is needed to match the observed 
neutrino luminosities \citep{Murase2020, Fiorillo2024b, Mbarek2024, Lemoine2025, Saurenhaus2025}, since it is this 
range that sets the available amount of power for the neutrino-producing TeV protons.
Predicting neutrino emission from kinetic plasma simulations requires extrapolations to
domains much larger than those accessible to present-day computing resources, which introduces a major uncertainty. 
In our simulations (Sec.~\ref{sec:simulations}), we find that ions develop broad nonthermal spectra, 
reaching the system-size (Hillas) limit of the simulation box, but the obtained power-law slopes are relatively steep 
with $p \gtrsim 3$. Macroscopically large domains, particularly those with turbulence driving on scales smaller than the 
system size, may present a more favorable situation, since in this case the typical 
ion escape time can become much longer than their acceleration time.
However, with larger amounts of energy accumulated in high-energy particles, 
the properties of the turbulent flow could be modified via 
the cosmic-ray feedback \citep{Lemoine2024}. Similarly as for the thermal particles, 
we may therefore expect that some form of self-regulation occurs also for the cosmic-ray 
component \citep[e.g., as suggested by][]{Lemoine2025}. 
In addition, nonthermal ion acceleration can be also facilitated by shocks 
\citep{Inoue2019,NhatLy2026}. In particular, 
acceleration by multiple colliding shocks \citep[e.g.,][]{Bykov2013} 
within a large turbulent volume could provide a possible mechanism for producing a hard ion power law. 
A more detailed investigation of these aspects requires dedicated simulations with larger boxes, which are left for the future.

\subsection{Comptonized Emission Spectrum}
\label{sec:emission}

\paragraph{X-ray spectrum}Low-energy seed photons introduced into the domain are Comptonized by 
the coronal electrons and form a power-law continuum with an X-ray 
photon index $\Gamma_{\!\rm x}$, defined via ${\rm d}N/{\rm d}E\propto E^{-\Gamma_{\!\rm x}}$.
The power law extends from the injection energy $E_0$ to the peak  at $E_{\rm peak}\sim 100$ keV.
In steady state,
the energy balance between escaping photons and the power received by turbulence reads
$(E_{\rm esc} - E_0)\dot{n}_{\rm ph} \simeq (1 - q_{\rm i})B_0^2 v_{\rm A}/ (8\pi s_0)$, where $\dot{n}_{\rm ph}$ is the rate 
of soft photon supply (equal to the rate of escape in steady state) and $E_{\rm esc}$ is the typical energy of 
escaping photons. This can be rearranged into
\begin{align}
    A -1 \simeq \frac{\sigma_{\rm e}(1-q_{\rm i})m_{\rm e}c^2}{E_0 \eta_{\rm ph}},
    \label{eq:A}
\end{align}
where $A = E_{\rm esc} /  E_0$ is the photon amplification factor and 
$\eta_{\rm ph} = \dot n_{\rm ph} S / (n_0 v_{\rm A})$
is an appropriately normalized photon injection rate. 
Eq.~\eqref{eq:A} can be alternatively written
as $A - 1 \simeq \ell / \ell_0$ \citep{Haardt1993}, 
where $\ell_0 = S^2 E_0 \dot n_{\rm ph} \sigma_{\rm T} /(m_{\rm e}c^3)$
is the soft photon compactness supplied to the corona, and the compactness $\ell$ due to
turbulent dissipation is estimated in 
Eq.~\eqref{eq:lrad1}. \citet{Beloborodov1999} obtained a convenient fitting formula of the
form $\Gamma_{\!\rm x}\simeq (7/3) (A - 1)^{-\delta}$. For AGNs the parameter $\delta \simeq 0.1$. 
We can then
estimate the expected spectral index of the Comptonized radiation as:
\begin{align}
    \Gamma_{\!\rm x} \simeq (7/3) \left[\sigma_{\rm e} (1-q_{\rm i})m_{\rm e}c^2/ (E_0 \eta_{\rm ph})\right]^{-0.1}.
    \label{eq:Gamma_X}
\end{align}

The peak of the spectral energy density can be identified with the mean energy of the 
coronal electrons. A crude estimate can be obtained by balancing the radiative cooling power with the
escaping luminosity. Neglecting Klein-Nishina corrections \citep{Moderski2005}, one can estimate
$\Theta_{\rm eff} \sim E_{\rm esc}/[4E_{\rm ph}\tau_{\rm T}(1+\tau_{\rm T})]$, 
where $E_{\rm ph}\sim E_{\rm esc}$ is the mean energy of photons 
contained in the source and 
$\Theta_{\rm eff} = \langle u^2/3\rangle$ is an \emph{effective} 
dimensionless electron temperature \citep{Groselj2024}, obtained by averaging the particle 4-velocity $u = \gamma v/c$. 
The expected peak of the Comptonized spectrum should be 
then near $E_{\rm peak} \sim m_{\rm e}c^2[(1 + 3\Theta_{\rm eff})^{1/2} - 1]$.

\paragraph{MeV tail}Beyond the peak near $\sim$ 100 keV, the spectrum may exhibit an MeV tail, which is shaped by nonthermal 
electrons \citep{Ghisellini1993,Zdziarski1993,Veledina2011,Zdziarski2017,Fabian2017}. Nonthermal features are most sensitive to the dissipative plasma physics, 
and therefore the MeV tail (or lack thereof) can provide important physical
constraints. The coronal electrons experience rapid radiative cooling on a timescale 
shorter than the dynamic time of the turbulent cascade \citep{Groselj2024}. A very 
rapid and efficient energization mechanism is required to generate a nonthermal electron tail under such 
conditions. 

In our simulations (Sec.~\ref{sec:simulations}) we find that electron acceleration
is enabled by collisionless reconnection at intense current sheets, which 
are embedded intermittently into the turbulent flow \citep{Comisso2019}.
Under the expected coronal conditions, reconnection of 
intense current sheets proceeds in the trans-relativistic and moderate-guide-field regime with $\sigma_{\rm e}\gg 1$, 
$\sigma_{\rm i}\sim 0.1$, and 
$\delta B \sim B_0$ \citep{Rowan2017,Werner2018,Ball2018,Comisso2024}.\footnote{Existing intermittency models of magnetized plasma turbulence imply that the most intense structures of various sizes roughly preserve the driving-scale 
fluctuation amplitude \citep[e.g.,][]{Chandran2015}. Thus, when $\delta B/B_0 \sim 1$ at the driving scale, 
it is reasonable to assume that the most intense current sheets feature a moderate guide field. Note that while the filling fraction of intense sheets may be small, the amount of plasma reprocessed by such sheets during 
their lifetime can amount to a sizable fraction of the volume \citep{Comisso2019}.} 
For these parameters, \citet{Comisso2024} reports a broken electron power law with a hard exponent $p \approx 1$ 
between the thermal (i.e., Maxwellian) peak of the distribution and the 
break at $E_{\rm break} \simeq 0.1 \sigma_{\rm e}m_{\rm e}c^2$, which is followed by a steeper 
power law with $p\approx 3.5$ up to some cutoff energy $E_{\rm max}$. 
An upper bound on $E_{\rm max}$ is provided 
by radiative cooling. By balancing the radiation drag with acceleration by the reconnecting 
electric field we find for $\tau_{\rm T}\sim 1$ that \citep[e.g., see][]{Werner2019,Sironi2020}:
\begin{align}
    E_{\rm max} \lesssim E_{\rm rad} \simeq m_{\rm e}c^2\!\left[\frac{\beta_{\rm rec} (v_{\rm A}/c)\,\sigma_{\rm e}^{1/2}s_{\rm esc}/d_{\rm e}}
    {f_{\rm KN}\ell}\right]^{1/2},
    \label{eq:gamma_rad1}
\end{align}
where $d_{\rm e} = [m_{\rm e}c^2 / (4\pi n_0 e^2)]^{1/2}$ is the electron skin depth, $\beta_{\rm rec}\simeq 0.1$ is the 
collisionless reconnection rate, and $f_{\rm KN}$ is a Klein-Nishina correction factor \citep{Mehlhaff2021}.
For a realistic size of an AGN corona (of the order of $\sim 10^{13} d_{\rm e}$ for a $10^7$ solar-mass black hole),
it can be easily checked that $E_{\rm rad} \gg \sigma_{\rm e}m_{\rm e}c^2$, and so the 
electrons can be accelerated to high energies, perhaps limited only by the size of the
(turbulent) reconnection layers \citep{Zhang2021}. The addition of synchrotron loses does not change this
conclusion. An expression equivalent to \eqref{eq:gamma_rad1} can be obtained
for synchrotron loses if one omits $f_{\rm KN}$ and replaces $\ell$ with the magnetic 
compactness $\ell_B\sim \tau_{\rm T}\sigma_{\rm e}$ \citep{Beloborodov2017}, which is of similar order as $\ell$ (by comparison with Eq.~\eqref{eq:lrad1}).

In a turbulent box, the expected average nonthermal spectrum is a 
composite between the one injected at current sheets and the one from electrons in the bulk of the turbulent volume. 
Outside the current sheets electrons are energized less efficiently, and so their box-averaged spectrum 
will be generally softer than the one produced at current sheets alone. In particular, it is 
reasonable to expect that the hard power-law range (below $E_{\rm break}$) 
may not show up clearly in the average spectrum. For additional results and discussions on this matter see 
Appendix~\ref{sec:injection}.

The photons can be in principle upscattered up to the maximum electron energy $E_{\rm max}$.
However, in the gamma-ray band the corona becomes opaque to 
pair production \citep{Murase2020}. The optical depth for absorption of gamma-rays can be estimated as
$\tau_{\gamma\gamma} \simeq \tau_{\rm T} n_1/(5n_0)$, where $n_1$ is the density 
of target photons with typical energy
$E_1 \simeq 2m_{\rm e}^2c^4/E_2$, for a given gamma-ray with energy $E_2$.
For target photons within the range of the X-ray power law, 
$n_1/n_0\simeq (n_{\rm ph}/n_0)(\Gamma_{\rm x}-1)(E_1/E_0)^{1-\Gamma_{\rm x}}$, where $n_{\rm ph}$ is the total
photon density.
The average ratio $n_{\rm ph}/n_0$ can be estimated from \eqref{eq:A} by noting that 
$\dot n_{\rm ph} \simeq n_{\rm ph} c / [(\tau_{\rm T}+1)s_{\rm esc}]$. This gives:
\begin{align}
    \frac{n_{\rm ph}}{n_0} \simeq \frac{\sigma_{\rm e}(v_{\rm A}/c)(1-q_{\rm i})(\tau_{\rm T}+1)m_{\rm e}c^2}
    {2E_0(A-1)}.
    \label{eq:nph}
\end{align}
For representative parameters, the ratio $n_{\rm ph}/n_0$ is a very large number, of the order of a few ten thousand.
This implies that the corona becomes opaque as soon as the gamma-rays can start pair producing with
target photons from the peak of the X-ray spectrum, right below the steep MeV tail.
A crude but relatively robust estimate for the absorbed gamma-ray photon range is thus 
given by $E_{\rm abs} \gtrsim 2m_{\rm e}^2c^4/ 
E_{\rm peak} \sim 5$ MeV.

\subsection{Range of Applicability}
\label{ref:applicability}

Our model assumes a strongly turbulent electron-ion corona with weak 
collisional coupling between ions and electrons. Frequent Coulomb collisions between ions and electrons 
or copious pair production in MeV-photon collisions can invalidate our assumptions.
In Appendix~\ref{sec:collisions} we estimate the relevant collision rates and find that the applicable $\sigma_{\rm i}$ range is
bounded from below by electron-ion collisions and from above by pair production. For representative parameters, 
we estimate that the turbulent corona can maintain a two-temperature state when $\sigma_{\rm i} \gtrsim 0.04$. According to
\eqref{eq:lrad1}, this corresponds to $\ell \gtrsim 10$. Thus, a two-temperature weakly collisional corona is roughly equivalent to 
a radiatively compact corona, according to our model. Regarding pair production, we estimate that significant 
pair enrichment occurs when $\sigma_{\rm i}\gtrsim 0.24$, which corresponds to the high 
end of the relevant range where $\ell \sim 100$ (see Eq.~\eqref{eq:sigma_range}).

\subsection{Alternative Dissipation Scenarios}

Alongside turbulence, dissipation in the corona can be mediated by shocks and/or reconnecting current sheets. 
We stress that our model does not exclude these alternative dissipation mechanisms, and in fact it combines elements
of both shocks and reconnecting sheets. A reconnection model
was developed by \citet{Beloborodov2017} and explored in a series of numerical 
simulations \citep{Sironi2020,Sridhar2021,Sridhar2023,Sridhar2025}. While current sheets can form naturally as part of the
turbulent cascade, our present model differs from the work by \citet{Beloborodov2017} in that
we consider a more volume-filling dissipation scenario, rather than an isolated reconnecting current sheet
in a quiescent background plasma. The latter would imply that dissipation takes place only in a small volume fraction of the
corona, which would modify our estimates. Furthermore, the dynamic timescale for the turbulence model is 
$\sim s_{\rm esc}/v_{\rm A}$, while in the model by \citet{Beloborodov2017} the dynamic timescale is longer by a 
factor of $\sim 1/\beta_{\rm rec}\sim 10$.

\section{Numerical Method}
\label{sec:method}

In support of our theoretical model, we perform radiative kinetic simulations of driven electron-ion plasma 
turbulence using the PIC code \textsc{Tristan-MP v2} \citep{tristanv2}.
Our numerical experiments model a local patch of the turbulent 
corona as an open system for photons and charged particles. Low-energy
particles (photons, electrons and ions), representing the influx from a 
cold medium in thermal equilibrium (e.g., a thin accretion disk), are continuously introduced into 
the domain. The inserted particles are then energized by the turbulent 
cascade and leave the box via diffusive escape.
The evolution of the radiation in the box and its feedback on the kinetic plasma
is determined via a Monte Carlo model of 
Compton scattering between the electrons and
photons \citep[e.g., see][]{DelGaudio2020,Groselj2024}.
The heavier ions are not coupled to the radiation; their steady state is thus obtained
by a balance between turbulent energization and
diffusive escape \citep{Gorbunov2025}. The latter is crucial for overcoming 
the limitations of closed domains, 
where ions heat up indefinitely \citep[e.g.,][]{Zhdankin2021}.

For computational convenience, we perform the simulations in a 2D 
periodic box of size $S\times S$.  While the spatial domain is 2D, the particle velocities and electromagnetic fields 
have components in all 3 directions (i.e., the simulations are 2D3V).
A mean magnetic field $\vec B_0 = B_0\hat{\vec z}$ is imposed in the out-of-plane $z$-direction. Turbulence is driven by applying 
an external time-varying electric current in the 
form of a ``Langevin antenna'' \citep{TenBarge2014}. 
The antenna is configured to drive 
large-amplitude magnetic fluctuations with $\delta B \sim B_0$
and polarization perpendicular to $\vec B_0$.
We drive the external current at modes with wavenumbers $(k_x S /2\pi,\,k_y S / 2\pi)  = (1,0),\,(0, 1),\,(1, 1),\,(1, -1)$ and 
frequencies $\omega_0 = \pm 1.3v_{\rm A}(2\pi/S)$.
The antenna decorrelation rate \citep{TenBarge2014} is $\gamma_0 = 0.5\omega_0$. 

The particles are inserted into the domain from an external medium at fixed temperature
$T_0$. The momenta of the inserted particles are sampled from a Maxwellian distribution
for ions and electrons, and from a Planck spectrum for the photons.
The thermal momenta of the charged particles 
are sampled in the local $E\times B$ fluid frame and then boosted into the simulation frame. 
In order to conserve charge, we insert new ions and electrons at any given time at the
exact locations where particles of the same charge just escaped from the box. 
The mean density $n_0$ of ions or electrons is thus fixed by construction.
For the seed photons, we prescribe a fixed volumetric injection rate, which is self-consistently balanced by diffusive escape once a steady state
has been reached. Following \citet{Groselj2024} and \citet{Gorbunov2025}, 
we implement escape by tracking the displacement of each
particle from its injection location. Particles escape when they diffuse over a
distance $s_{\rm esc} = S/2$ in $x$ or $y$, which mimics escape from a turbulent
accelerator of size $S$.

We perform a set of 2D simulations at a reduced ion-electron 
mass ratio of $m_{\rm i}/m_{\rm e} = 144$ for different values of the ion
 magnetization $\sigma_{\rm i} = 0.035,\,0.1,\,0.19$. The electron magnetization 
 $\sigma_{\rm e} = \sigma_{\rm i}m_{\rm i}/m_{\rm e}$ is well above unity, as expected for 
 black-hole coronae (see Sec.~\ref{sec:turb_theory}).
The electron skin depth in units of the cell size is $d_{\rm e} = 2.5\Delta x$.
Compton scattering is resolved on spatial tiles spanning $15$ cells of the PIC grid in each dimension.
The time step of the PIC scheme and of the Compton scattering
algorithm is $\Delta t = 0.29 \Delta x / c$.
Our square domain has a linear size $S = 256\,d_{\rm i}$ in runs with $\sigma_{\rm i} = 0.035,\,0.1$ 
and $S = 448\,d_{\rm i}$ in the run with $\sigma_{\rm i} = 0.19$, where $d_{\rm i} = (m_{\rm i}/m_{\rm e})^{1/2}d_{\rm e}$
is the ion skin depth. This amounts to a numerical resolution of 7,680$^2$ in runs with $\sigma_{\rm i} = 0.035,\,0.1$
and 13,440$^2$ in the $\sigma_{\rm i} = 0.19$ run.
In all simulations, the average steady-state number of photons per cell (of the PIC grid) is roughly 100.
Ions and electrons are each represented with 
$10$ particles per cell for $\sigma_{\rm i} = 0.035,\,0.1$ and with $5$ particles per cell
for $\sigma_{\rm i} = 0.19$. The external medium in thermal equilibrium, from which we sample 
the newly inserted particles, has a temperature $T_0 = 10^{-5} m_{\rm e}c^2$. The corresponding mean energy
of the blackbody seed photons is $E_0 \approx 2.7\!\!\;\times\!\!\;10^{-5}m_{\rm e}c^2$.
The reference Thomson optical depth is $\tau_{\rm T} = \sigma_{\rm T} n_0 s_{\rm esc} = 1$.
For each run, we choose a suitable soft-photon injection rate $\eta_{\rm ph} = \dot n_{\rm ph} S / (n_0 v_{\rm A})$,
so that the obtained X-ray spectral index is in good agreement with 
observations (see Eq.~\eqref{eq:Gamma_X}). To this end, we performed a series of 
preparatory test runs, until an appropriate choice of $\eta_{\rm ph}$ 
was found for a given $\sigma_{\rm i}$. In our production runs, we use 
$\eta_{\rm ph} = 3.18\;\!\!\times\;\!\!10^3,\, 9.07\!\!\;\times\!\!\;10^3,\, 2.10\!\!\;\times\!\!\;10^4$
for $\sigma_{\rm i} = 0.035,\,0.1,\,0.19$, respectively.

The predicted spectra are compared with X-ray observations of NGC 4151 from 
Swift/BAT, INTEGRAL, and NuSTAR. The Swift/BAT and INTEGRAL data represent long-term
averages over several years of observations. The NuSTAR data are from Nov 14, 2012. The PIC simulation spectra are 
corrected for absorption and reflection when comparing with observations. We adopt a simple model, consistent
with an earlier study of the same source \citep{Hinkle2021}.
For the reflection component, use the \textsc{\mbox{reflect}} model \citep{Magdziarz1995} of 
\textsc{xspec} \citep{XSPEC} with a reflection factor $R=0.494$ and inclination angle $i = 45^\circ$. The spectra are additionally corrected for 
absorption using the \textsc{\mbox{ztbabs}} model of \textsc{xspec}, with hydrogen column density $N_{\rm H} = 10.3\!\!\;\times\!\!\; 10^{22} / {\rm cm}^2$,
redshift $z = 0.003$, and elemental abundances from \citet{Wilms00}.

\section{Simulation Results}
\label{sec:simulations}

Below we present results from our set of local PIC simulations of strongly turbulent electron-ion coronae of accreting black holes.
The simulations are performed for different strengths of the ion magnetization $\sigma_{\rm i}$, which correspond to different values of the 
steady-state radiative compactness $\ell$. In Sec.~\ref{sec:turbulence} we investigate the properties of the turbulent flow, which is 
followed by the analysis of charged particle and photon spectra in Sec.~\ref{sec:acceleration}.

\subsection{Properties of the Turbulent Flow}
\label{sec:turbulence}

\begin{figure}[htb!]
\centering
\includegraphics[width=\columnwidth]{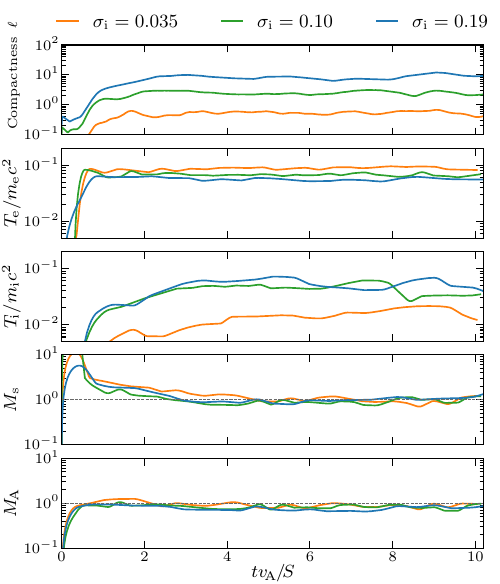}
\caption{\label{fig:traces} Time evolution and approach to steady state in our radiative PIC simulations of 
turbulence with ion magnetizations $\sigma_{\rm i} = 0.035,\, 0.1,\, 0.19$. Shown from top to bottom are the 
box-averaged compactness $\ell$, proper electron kinetic temperature $T_{\rm e}$, proper ion 
kinetic temperature $T_{\rm i}$, the turbulent sonic Mach number $M_{\rm s}$, and the Alfv\' enic Mach number $M_{\rm A}$.}
\end{figure}

\paragraph{Global evolution}Let us first demonstrate how our simulations 
settle into a trans-sonic and trans-Alfv\' enic steady state, consistent 
our theoretical arguments (Sec.~\ref{sec:ion_heating}). Fig.~\ref{fig:traces} shows the time
evolution of various box-averaged quantities.
A quasi-steady state is attained in roughly 4 Alfv\' en crossing times $S/v_{\rm A}$.
In the bottom two panels we show the evolution of the sonic Mach number $M_{\rm s} = v_{\rm rms}/c_{\rm s}$ 
and of the Alfv\' enic Mach number $M_{\rm A} = v_{\rm rms}/v_{\rm A}$, where $v_{\rm rms}$ is the root-mean-square ion bulk
velocity and $c_{\rm s} = (\Gamma_{\rm i} T_{\rm i}/m_{\rm i})^{1/2}$ is the 
ion sound speed (using $\Gamma_{\rm i}=5/3$ for the adiabatic index). We find that the steady state is characterized by $M_{\rm A}\sim M_{\rm s}\sim 1$.
For the considered range of ion magnetizations ($\sigma_{\rm i}\lesssim 0.2$), 
the corresponding steady-state proper ion kinetic temperature $T_{\rm i}$ is non-relativistic 
or very mildly relativistic, as anticipated in Eq.~\eqref{eq:Ti}. The box-averaged proper electron kinetic temperature 
is of the order of $T_{\rm e}\sim 0.1 m_{\rm e}c^2$, and shows a mild 
decline with increasing $\sigma_{\rm i}$. The latter can be attributed to the increased cooling strength at 
larger values of the compactness \citep{Groselj2024}, which scales as $\ell \sim \sigma_{\rm e}(v_{\rm A}/c)\sim(m_{\rm i}/m_{\rm e})\sigma_{\rm i}^{3/2}$ 
(see Eq.~\eqref{eq:lrad1}). In the top panel of Fig.~\ref{fig:traces} we calculate $\ell = S^2\dot{U}_{\rm esc}\tau_{\rm T}/(s_{\rm esc}n_0 m_{\rm e}c^3 )$ 
by measuring the box-averaged power of escaping photons per unit volume. In steady state, 
we find $\ell \approx 0.6,\, 2.5,\, 8.5$ in simulations with
$\sigma_{\rm i} = 0.035,\, 0.1,\, 0.19$, respectively. The run with $\sigma_{\rm i} = 0.19$ is therefore most representative 
of the high-compactness regime. Note that the obtained $\ell$ values are lower than those expected for a proton-electron plasma 
due to our use of a reduced ion-electron mass ratio ($m_{\rm i}/m_{\rm e} = 144$), which is a choice imposed by computational constraints.

\begin{figure*}[htb!]
\centering
\includegraphics[width=\textwidth]{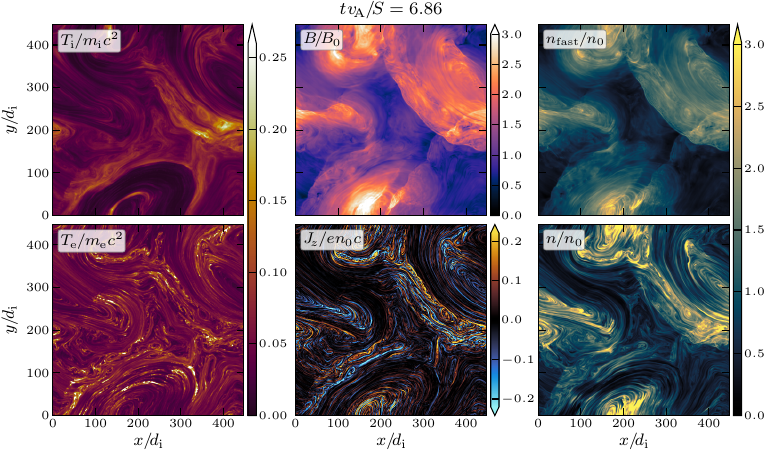}
\caption{\label{fig:visuals} Visualization of turbulent fields in 
our simulation with $\sigma_{\rm i} = 0.19$ at time $t = 6.86\, S/v_{\rm A}$. The
left panels show the proper ion and electron kinetic temperatures ($T_{\rm i}$ and $T_{\rm e}$). 
In the middle panels we show the magnetic field magnitude $B$ and the out-of-plane electric current $J_z$.
Finally, the right panels depict the ion density $n$ and the 
density of fast-mode fluctuations $n_{\rm fast}$. For easier comparison with the total density $n$, 
we include the mean value in the visualization of the fast mode (i.e., $n_{\rm fast} = \delta n_{\rm fast} + n_0$).
An animated version of this figure is available online at \url{https://youtu.be/0W-b242WMhw}. The animation
lasts 50 s and shows the spatiotemporal 
evolution of the turbulent fields from the start ($t v_{\rm A}/S=0$)
to the end ($t v_{\rm A}/S = 10.3$) of the simulation, in the same format as the
static figure.}
\end{figure*}

\paragraph{Spatial structure}In Fig.~\ref{fig:visuals} we depict the spatial structure of turbulent fluctuations in the quasi-steady state of the simulation
with $\sigma_{\rm i} = 0.19$. It can be seen that the plasma develops a two-temperature state with $T_{\rm i} \gg T_{\rm e}$. 
The turbulent field is populated with intense current sheets and magnetized shocks, which energize the particles.
By means of an MHD mode decomposition (see Appendix~\ref{sec:modes}), we extract the density fluctuations corresponding to 
fast modes and find that shocks form via steepening of large-amplitude fast waves. The presence of fast modes is a
natural consequence of our turbulence driving with 
$\delta B/B_0 \sim 1$.\footnote{A second-order expansion of the magnetic field magnitude 
gives $B/B_0 \approx 1 + \delta B_z/B_0  + (\delta B_{\perp}/B_0)^2/2$, where $\delta B_z$ and $\delta B_\perp$ are the fluctuations 
parallel and perpendicular to $\vec B_0$, respectively. Thus, even though our forcing induces Alfv\' enic-like 
fluctuations with $\delta B\approx \delta B_\perp$, at large amplitudes a portion of the injected power is converted into 
compressible fast modes via magnetic-pressure fluctuations.} 
Unlike the rest of the turbulent fluctuations (e.g., the plasma density $n$ in Fig.~\ref{fig:visuals} modulo the
fast mode contribution $n_{\rm fast}$, or the $J_z$ fluctuations), which develop a filamentary structure typical for Alfv\' enic 
turbulence, the fast modes feature a combination of relatively smooth large-scale structures and shock discontinuities (seen most clearly 
in snapshots of $B$ and $n_{\rm fast}$). Snapshots of $B$ also show that shocks tend to spread into regions 
with below-average magnetic field magnitude ($B < B_0$), where the local 
Alfv\' en speed is lower than the reference value $v_{\rm A}\simeq c\sigma_{\rm i}^{1/2}$. This allows for formation of shocks
with Alfv\' enic Mach numbers up to $\sim$ a few. For a more detailed analysis of shock properties see Appendix~\ref{sec:injection}.

Figure~\ref{fig:visuals} also shows that the fluctuations in the ion and electron temperatures are correlated with the
turbulent structures. In particular, the electron kinetic temperature is well correlated with 
electric current sheets. On the other hand, the ion temperature is enhanced at both shocks and current sheets, but the apparent 
correlation is weaker than for electrons. This is because the ions do not experience the rapid radiative cooling, and so
their temperature fluctuations can spread over large regions of the volume away from the local energization sites.
In Appendix~\ref{sec:injection} we analyze the local particle distributions near current sheets and shocks and find that 
electrons are injected into nonthermal distributions predominantly at current sheets, whereas the ions can be injected at both 
shocks and current sheets. We also find that the nonthermal electrons develop anisotropic distributions with 
small pitch angles with respect to the local magnetic 
field \citep[on this note, see also][]{Comisso2024,Vega2024}.

\begin{figure}[htb!]
\centering
\includegraphics[width=\columnwidth]{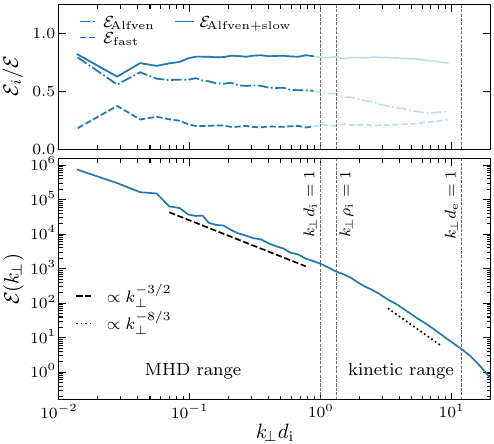}
\caption{\label{fig:turb_spectra} Turbulent energy spectrum in our 
simulation with $\sigma_{\rm i}=0.19$, averaged over the steady state from $t = 4\,S/v_{\rm A}$ until the
end of the run at $t = 10.3\,S/v_{\rm A}$. In the bottom panel we show the 1D energy spectrum ${\mathcal E}(k_\perp)$ as 
a function of the perpendicular wavenumber $k_\perp$. Power-law slopes are shown for reference.
In the top panel we show the relative energy 
content of fluctuations identified as the MHD Alfv\'en, slow, and fast mode (see Appendix~\ref{sec:modes} for details).}
\end{figure}

\paragraph{Spectrum of turbulence}In Fig.~\ref{fig:turb_spectra} we analyze the 1D turbulent energy spectrum ${\mathcal E}(k_\perp)$ 
as a function of the wavenumber $k_\perp$ perpendicular
to $\vec B_0$.  The total turbulent energy is obtained here as the sum of energy from the electromagnetic,
bulk kinetic, and density fluctuations (see Appendix~\ref{sec:modes} for details).
Over the MHD scale range ($k_\perp \lesssim 1/d_{\rm i}$) the turbulence spectrum shown in the bottom panel of Fig.~\ref{fig:turb_spectra} 
forms a power law with a slope 
close to $-3/2$. This is consistent with refined models of MHD turbulence that incorporate the dynamic alignment effect \citep{Boldyrev2006}.
Near the transition into the kinetic range the spectrum then steepens as a result of 
dispersive wave physics and collisionless damping of the turbulent cascade \citep{Howes2015a}. 
The position of the spectral break is more closely aligned with the scale of the 
ion thermal gyroradius $\rho_{\rm i} = \sqrt{\beta_{\rm i}}\, d_{\rm i}$ (estimated from the steady-state ion 
beta $\beta_{\rm i}\approx 0.57$), 
rather than with $d_{\rm i}$, in accordance with theoretical expectations 
for low-frequency Alfv\' enic turbulence \citep{Schekochihin2009}. The position of the break could be also affected by
shocks, which have a typical width of a few ion skin depths (Appendix~\ref{sec:injection}).
From the steady-state spectrum we also estimate the
turbulence correlation length, which can be taken as a proxy for the driving scale $s_0$.
We find $s_0 \simeq \pi\left(\int k_{\perp}^{\;\!-1}{\mathcal{E}}(k_\perp){\rm d}k_\perp\right)/\left(\int {\mathcal{E}}(k_\perp){\rm d}k_\perp\right)\approx S/3$, 
and so the effective ratio $s_{\rm esc} / s_0 \approx 3/2$ is slightly larger than unity for our choice of the forcing.

3D radiative PIC simulations of pair plasmas by \citet{Groselj2024} demonstrated
that turbulent bulk motions can be subject to radiative damping, which
gives rise to steeper (non-universal) turbulence spectra over the MHD scale range. On the other hand, the spectrum shown in Fig.~\ref{fig:turb_spectra}
does not show any apparent signs of abnormal steepening, and is instead reminiscent of turbulent 
cascades in more conventional weakly radiating plasmas, such as the solar wind \citep{Kiyani2015}. 
To put our results in the context of the previous work by \citet{Groselj2024}, we estimate the amount of power passed to the 
photons via bulk Comptonization. Presently we are unaware of a generally accepted definition of bulk Comptonization 
in a kinetic electron-ion plasma. Here, we employ a definition 
based on the transfer of bulk momentum between electrons and photons, which seems like natural choice from an energetics point of view.
At each time step, we compute the local net change of the photon momentum $\delta\vec P_{\!\rm tile}$, and from
there we estimate $\Delta E_{\rm bulk} = \sum c\left(|\vec P_{\!\rm tile} + \delta\vec P_{\!\rm tile}| - |\vec P_{\!\rm tile}|\right)$, where the sum runs over the 
spatial tiles of our Monte Carlo Compton 
scattering scheme (see Sec.~\ref{sec:method}), and $\vec P_{\rm tile}$ is the net photon momentum in the tile before scattering. Each tile 
contains of the order of $\sim 10^4$ computational photons and $\sim 10^3$ electrons, so that $\delta \vec P_{\!\rm tile}$ is a measure of the local bulk momentum
transfer between the electron and photon gas. The power lost via bulk Comptonization is then obtained by 
averaging $\Delta E_{\rm bulk}/\Delta t$ over steady state. This gives for the bulk Comptonization fraction 
$f_{\rm bulk} \approx 0.18,\, 0.37,\, 0.51$ in simulations with $\sigma_{\rm i} = 0.035,\, 0.1,\, 0.19$, respectively. 
That $f_{\rm bulk}$ increases with $\sigma_{\rm i}$ is understandable, since the magnetization sets the Alfv\' en speed, and in turn the
typical magnitude of the turbulent bulk motions. This suggests that the turbulent motions in an electron-ion corona with $\sigma_{\rm i}\sim 0.1$ 
may not be sufficiently relativistic to give rise to a significant radiative steepening of the turbulence spectrum. In this regard, it may seem
surprising that the measured values of $f_{\rm bulk}$ are still relatively high.
The bulk momentum of a (mildly) relativistic 
fluid is $w\Gamma_{\rm bulk}^2\vec v_{\rm bulk}/c^2$, where $w$ is the proper enthalpy density,
$\vec v_{\rm bulk}$ is the bulk velocity and $\Gamma_{\rm bulk} = \left[1 - (v_{\rm bulk}/c)^2\right]^{-1/2}$.
Thus, the amount of power lost via bulk Comptonization can be enhanced by mildly relativistic electron
temperatures (via the enthalpy term), according to our procedure for estimating $f_{\rm bulk}$. This may be a reason why the 
estimated bulk Comptonization fractions are relatively high.

To further analyze the nature of turbulent fluctuations, we 
decompose in the top panel of Fig.~\ref{fig:turb_spectra} the spectra into contributions from the 
MHD Alfv\' en, slow, and fast modes (see Appendix~\ref{sec:modes} for details). At the largest scales ($k_\perp \lesssim 0.05 / d_{\rm i}$), fast modes
contain roughly 30\% of the total energy, less than 10\% is contained in slow modes, 
and the rest is carried by Alfv\' en modes.
The sum of fluctuations identified as either the Alfv\' en or slow mode is relatively
stable over the MHD scale range ($k_\perp \lesssim 1/d_{\rm i}$), in contrast to the fluctuations identified as Alfv\' en waves only, 
which feature an apparent decline with growing $k_\perp$. According to our classification scheme (see relation \eqref{eq:slow} in Appendix \ref{sec:modes}), 
any pressure balanced fluctuation is categorized as the slow mode. However, in a kinetic plasma 
Alfv\' en waves gradually develop a finite compressibility as their wavelength approaches the 
ion kinetic scales \citep{Howes2006}. In addition, slow modes are not expected to dynamically
couple to the Alfv\' enic cascade and are instead merely passively mixed by the Alfv\' enic 
turbulence \citep{Schekochihin2009}. This suggests that the declining Alfv\' en mode fraction with increasing
$k_\perp$ can be largely attributed to the conversion of the Alfv\' en modes into compressible \emph{kinetic} Alfv\' en waves. Therefore, 
the sum $\mathcal{E}_{\rm Alfven + slow}$ may be a better representation of the true energy content of the 
Alfv\' enic cascade, while the true energy content of slow modes is likely minor.

It is known that the cascade of turbulent fast modes is isotropic in full 3D geometry \citep{Cho2003,Takamoto2016}, 
in contrast to the Alfv\' enic cascade which is anisotropic in wavenumber space \citep{Goldreich1995}. 
Therefore, 3D kinetic simulations are required for a complete picture of how the partitioning among different modes 
shapes the nature of the turbulent cascade and the resulting particle energization. An exact 
3D equivalent of our 2D PIC simulations is computationally infeasible at present. 3D aspects of the problem
could be instead studied with reduced-kinetic or fluid-like models, provided that such approximations are 
appropriate for the specific question at hand.  This direction is left for future work.

\subsection{Energy Partitioning, Nonthermal Acceleration and Emission Spectrum}
\label{sec:acceleration}

\begin{figure}[htb!]
\centering
\includegraphics[width=\columnwidth]{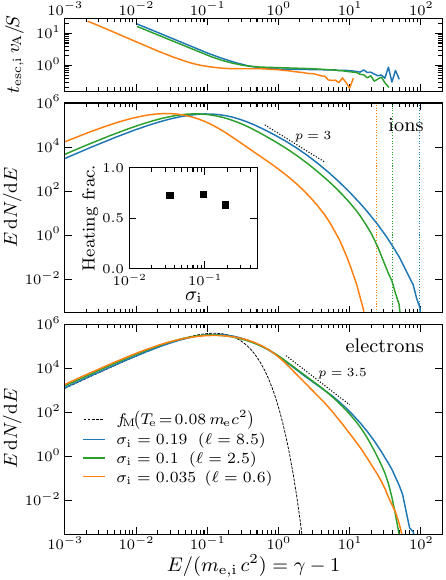}
\caption{\label{fig:prtl_spectra} Steady-state particle spectra in our simulations with different 
strengths of the ion magnetization $\sigma_{\rm i}$. The middle panel shows the ion energy spectra 
and the bottom panel the electron spectra. Energy is measured in units of the particles' own
rest mass ($m_{\rm i}c^2$ for ions and $m_{\rm e}c^2$ for electrons).
Power-law slopes are indicated with black dotted lines for reference. In the bottom panel we fit a Maxwellian 
distribution to the low-energy part of the spectrum (dashed black curve). The top panel shows the ion escape time $t_{\rm esc,i}$. Finally, the inset 
shows the ion heating fraction.}
\end{figure}

\paragraph{Ion and electron spectra}We now analyze the distribution of particles 
energized by the turbulent cascade. The steady-state distributions are shown in Fig.~\ref{fig:prtl_spectra}.
The ion spectra are shown in the middle panel and the electron spectra in the bottom panel. Both the ion and the 
electron distributions feature a 
quasi-thermal (i.e., Maxwellian) part and a high-energy nonthermal tail. In accordance with Eq.~\eqref{eq:Ti},
the thermal peak of the ion distribution shifts to higher energies at larger values of $\sigma_{\rm i}$, which correspond 
to larger values of the radiative compactness $\ell$. On the other hand, the position of the electron
thermal peak is independent of $\sigma_{\rm i}$ (or $\ell$), which shows that the effective electron temperature is 
primarily set by the optical depth (equal to $\tau_{\rm T}=1$ in all runs), as explained in Sec.~\ref{sec:emission}.

Figure~\ref{fig:prtl_spectra} also shows that higher magnetizations lead 
to more pronounced nonthermal features.  Nonthermal ions with energies greater than 4 times the mean of their distribution contain roughly 1/5 of the total
kinetic energy for $\ell\approx 0.6$ and about 1/4 for $\ell \approx 2.5,\,8.5$.
With the exception of the low-compactness simulation ($\ell\approx 0.6$), the nonthermal ions are accelerated up to the maximum energy set 
by the size of the simulation box $E_{\rm max} = m_{\rm i}c^2\sigma_{\rm i}^{1/2} s_{\rm esc}/d_{\rm i}$ (dotted vertical lines in 
the middle panel of Fig.~\ref{fig:prtl_spectra}), where the particle gyroradius $\rho_{\max} = s_{\rm esc}$. The local power-law index
in the nonthermal parts of the ion distribution is $p\gtrsim 3$. The electron power-law index hardens 
with growing magnetization (or compactness). For $\ell > 1$,
we find $p\approx 3.5$ for electrons. This is broadly consistent with our 
theoretical discussion from Sec.~\ref{sec:emission}, where we argue that
electrons are predominantly accelerated at intense reconnecting current sheets, in which case a power-law index 
not far from $p\approx 3.5$ is expected for $\sigma_{\rm i}\sim 0.1$ 
and $\delta B/B_0\sim 1$ \citep{Comisso2024}. It can be also noticed 
that in between the electron thermal peak and the
power-law tail lies a suprathermal range, where the spectrum is relatively flat.
This could be in part related to the hard power-law range expected 
for high-$\sigma_{\rm e}$ reconnection (see Sec.~\ref{sec:emission}), but in a turbulent volume the hard 
power law could be obscured by contributions from particles outside the current sheets, where the spectrum is softer.
For a more detailed analysis of particle injection into nonthermal populations near shocks and current sheets see 
Appendix~\ref{sec:injection}.

\paragraph{Ion escape time}In the top panel of Fig.~\ref{fig:prtl_spectra} we calculate the energy-dependent 
ion escape time $t_{\rm esc,i} = f_{\rm i}(E) / \dot f_{\rm esc,i}(E)$, where $\dot f_{\rm esc,i}(E)$ is the 
measured flux of escaping ions per unit energy and $f_{\rm i}(E) = {\rm d}N_{\rm i}/{\rm d}E$ is the distribution 
of ions contained in the box \citep[see also][]{Gorbunov2025}. At energies above the thermal peak, the measured 
escape time is of the order of an Alfv\' en crossing time $S/v_{\rm A}$. This shows that even moderate-energy particles are not well confined by the
source, when the turbulence is strongly driven on a scale comparable to the system size. An alternative energy-independent measure may be obtained by
calculating the average lifetime or confinement time of particles in the box, from the moment they are introduced into the domain and until they
escape. In steady state, we measure an average confinement time of roughly $2 S/v_{\rm A}$, which is in reasonable agreement with the energy-dependent
measure shown in the top panel of Fig.~\ref{fig:prtl_spectra}.

\paragraph{Ion heating fraction}The inset plot of Fig.~\ref{fig:prtl_spectra} shows the relative amount of turbulence power deposited into the
escaping ions. We calculate the ion fraction as $q_{\rm i} = (L_{\rm i} - L_{\rm i0})/\sum_s(L_{s}-L_{s0})$,
where $L_{ s}$ is the average escaping luminosity of a given species ($ s=$ ions, electrons, or photons), and $L_{s0}$ is the average power from new particles
supplied into the box (essentially, $L_{s} - L_{s0}$ is the average power species $s$ receives from the turbulent cascade).
We measure $q_{\rm i} \approx 0.72,\, 0.73,\, 0.63$ in simulations with $\sigma_{\rm i} = 0.035,\, 0.1,\, 0.19$, respectively. 
This is broadly consistent with our theoretical estimate, 
which predicts $q_{\rm i}\simeq 0.5$ (Eq.~\eqref{eq:qi}). Our theoretical argument is rather generic and does not specify how exactly the ions 
are energized in the steady state. More detailed heating predictions could be built
upon specific energization mechanisms. For representative parameters of the steady state 
($\sigma_{\rm i}\sim 0.1,\, \beta_{\rm i}\sim 1,\, \delta B/B_0 \sim 1,\ T_{\rm i}\gg T_{\rm e}$),
heating prescriptions based on either turbulent
wave-particle interactions \citep{Howes2010a,Kawazura2019,Kawazura2020,Adkins2025} or 
magnetic reconnection \citep{Rowan2017,Werner2018,Rowan2019} would be compatible with values close to $q_{\rm i}\sim 0.5$, 
while heating by low Mach number shocks would imply $(1-q_{\rm i})\ll 1$ \citep{Sironi2024}.
We emphasize that $q_{\rm i}$ measures the total fraction of energy
deposited into the escaping ions, in the form of thermal or nonthermal particles. In other words, $q_{\rm i}$ can be taken as an upper limit for
the available fraction of energy that can be deposited into cosmic-ray ions. Finally, it is worth 
mentioning that our lowest magnetization run with $\sigma_{\rm i} = 0.035$ would be realistically in the regime with significant collisional coupling
between ions and electrons (see Appendix~\ref{sec:collisions}). The $\sigma_{\rm i} = 0.035$ run is included here to better illustrate the general 
dependence on $\sigma_{\rm i}$ in the absence of ion-electron collisions, but caution is advised when applying the low-$\sigma_{\rm i}$
results to real sources, where the collisional coupling could play a role.

\begin{figure}[htb!]
\centering
\includegraphics[width=\columnwidth]{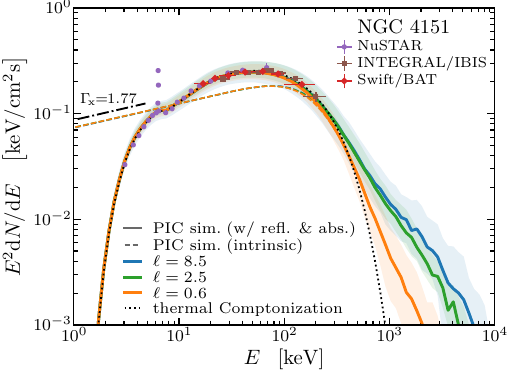}
\caption{\label{fig:fits} Comparison of predicted emission spectra with X-ray observations of NGC 4151. Dashed curves show the 
intrinsic steady-state spectra obtained directly from our PIC simulations. Solid color curves show the PIC spectra corrected for 
absorption and reflection. For $E\lesssim 100$ keV, the PIC spectra 
obtained for different values of the compactness $\ell$ lie on top of each other and are thus nearly indistinguishable.
Shaded bands indicate the ranges over which the simulated spectra vary during 
the averaging interval from $t \approx 4 S/v_{\rm A}$ to $t \approx 10 S/v_{\rm A}$. The dotted black curve is the \textsc{\mbox{nthcomp}}
thermal Comptonization model of \textsc{xspec}, fitted to the X-ray continuum and corrected for absorption and reflection.}
\end{figure}

\paragraph{Emission spectrum}Figure~\ref{fig:fits} shows the predicted emission spectra from our model and compares the results with
observations of NGC 4151 from \mbox{NuSTAR}, \mbox{Swift/BAT}, and \mbox{INTEGRAL}. We take the \mbox{NuSTAR} measurements as a reference and normalize 
the other data by matching to the \mbox{NuSTAR} luminosity in the 26.2 to 68.2 keV band (where \mbox{NuSTAR} spectra overlap with the \mbox{INTEGRAL} 
and \mbox{Swift} results).
The intrinsic steady-state spectra of escaping photons are shown with dashed curves. The solid curves show the PIC simulation spectra corrected for
reflection and absorption (see Sec.~\ref{sec:method} for details).

Our simulated spectra are in excellent agreement with observations of NGC 4151. This demonstrates that our kinetic simulations probe 
the observationally relevant regime of turbulence, and so the obtained physical insight (e.g., concerning the energy partitioning) 
can be put into the context of real sources. The intrinsic X-ray photon index is close to 
$\Gamma_{\rm x} = 1.77$ (dash-dotted black line in Fig.~\ref{fig:fits}), which is in reasonable agreement 
with previous studies of NGC 4151 \citep[e.g., by][who found $\Gamma_{\rm x}=1.76$]{Lubinski2016}. For our choice of model
parameters (see Sec.~\ref{sec:method}) and for the measured values of $q_{\rm i}$, Eq.~\eqref{eq:Gamma_X} predicts $\Gamma_{\rm x} \approx 1.76,\, 1.77,\, 1.75$
in runs with $\sigma_{\rm i} = 0.035,\, 0.1,\, 0.19$, respectively. The peak of the (intrinsic) spectrum is near 80 keV, which 
is as well in agreement with our theoretical estimate 
$E_{\rm peak} \sim m_{\rm e}c^2\left\{[1 + 3/(4\tau_{\rm T}(1\!\!\;+\!\!\;\tau_{\rm T}))]^{1/2} - 1\right\}\sim 90$ keV.
Thus, our numerical results are well aligned with theoretical expectations for the model.

Figure~\ref{fig:fits} also shows that the predicted X-ray spectrum is degenerate, in the sense that simulations
employing different values of the compactness $\ell$ can be matched to observations. However, the differences become apparent above the peak of the 
spectrum, where larger values of the radiative compactness lead to a more pronounced MeV tail. As discussed in Sec.~\ref{sec:emission},
the MeV tail is shaped by nonthermal electrons, which are sensitive to the microphysical details of the dissipation. 
 The MeV tail generated by nonthermal electrons differs from predictions based on thermal Comptonization models, 
which predict a sharp cutoff above peak (see black dotted curve in Fig.~\ref{fig:fits}).
Larger values of the compactness imply higher electron magnetizations, which give rise to harder nonthermal tails from
electrons accelerated at reconnecting current sheets. Moreover, 
in the range around 1 MeV the spectrum is nearly unaffected by the (parameter-dependent) reflection modeling. Thus, future measurements
with MeV-band instruments \citep{Tomsick2024} can provide important physical constraints on the emission mechanism. Past attempts to
measure the MeV tail in NGC 4151 \citep{Zdziarski1996} suggest that the spectrum indeed extends into the MeV range, in broad agreement with
our predictions for $\ell > 1$.

\section{Discussion and Conclusions}

We developed a local emission model of black-hole coronae based on dissipation in a strongly turbulent electron-ion plasma.
Our theoretical estimates (Sec.~\ref{sec:theory}) are informed by a set of radiative 
PIC simulations (Sec.~\ref{sec:simulations}), which include self-consistent Compton scattering, as 
well as injection and diffusive escape of charged particles and photons. The simulations allow us to track from first principles the
turbulent partitioning of energy among ions, electrons, and photons. We show that the turbulent cascade regulates itself into 
a trans-sonic state, which features sporadic formation of magnetized fast-mode shocks (Sec.~\ref{sec:turbulence}). In steady state, 
the ion kinetic temperature far exceeds the electron kinetic temperature ($T_{\rm i}\gg T_{\rm e}$). 
Complementary to alternative arguments, based e.g.~on assuming that the ions are near the virial temperature \citep[e.g.,][]{Bambic2024a}, we interpret 
the two-temperature state as a natural outcome of the interplay between the dissipative microphysics and radiation.
In order to maintain the expected steady-state ion kinetic temperature, an order-unity fraction of the available power needs to be deposited
into the ions (see Sec.~\ref{sec:ion_heating}). In our simulations, we measure $0.6 \lesssim q_{\rm i} \lesssim 0.7$ for the ion heating fraction, 
which shows that most of the dissipated power is deposited into the ions.

The predicted emission spectra obtained from our PIC simulations are confronted with X-ray observations of NGC 4151, demonstrating
excellent agreement (Sec.~\ref{sec:acceleration}). We show that the MeV tail, which is presently poorly constrained by observations, can provide important 
insight into the nonthermal physics of the coronal electrons. In particular, we find that higher values of the radiative compactness $\ell$
give rise to a more pronounced MeV tail, which is shaped by nonthermal electrons. Our predictions can be 
tested with future MeV-band instruments
such as COSI \citep{Tomsick2024}. Presently, the diffuse cosmic X-ray background gives an 
upper limit on how strong on average the MeV tail generated by the nonthermal 
electrons might be. Measurements of the 
X-ray background indeed show an excess 
over conventional models in the $\sim$ 150 -- 1000 keV band \citep[e.g., see Figure 9 from][]{Marcotulli2022}.
In addition, MeV photons can be also generated in electromagnetic cascades of high-energy gamma-rays, 
born in hadronic reaction chains \citep[e.g.,][]{Murase2020}. Modeling of electromagnetic cascades 
is beyond the scope of the present work.

While our model gives an apparent good match to the observed X-ray spectrum of a bright AGN, it falls short of being able to predict the
emission in lower-energy bands, such as the UV/optical \citep[e.g.,][]{Steffen2006,Duras2020} or the 
mm band \citep[e.g.,][]{Kawamuro2022,Ricci2023}, and their connection to the X-ray luminosity. 
We stress that our local model only tracks
the soft photons intercepted by the hot corona. Realistically, only a fraction of the UV/optical 
photons from the disk is actually intercepted by the corona \citep{Done2007}. Similarly, the recently observed 
mm emission may come from a large-scale outflow launched from the corona \citep{Hankla2025}. Thus, accurate multi-wavelength predictions 
require a global model for the extended accretion disk and its corona \citep[e.g.,][]{Nagele2026}.

Finally, we find that a strongly turbulent corona can efficiently 
accelerate nonthermal particles (Sec.~\ref{sec:acceleration}). We investigate the origins of nonthermal acceleration and
show that the electrons are preferentially injected into nonthermal power laws at intense current sheets, while the ions can be injected at 
both shocks and current sheets (Appendix~\ref{sec:injection}). In our limited-size simulation boxes, the nonthermal ions can be accelerated 
up to the system size (Hillas) limit, albeit with a relatively steep spectrum (with power-law index $p\gtrsim 3$). Simulations with significantly
larger boxes are required to confidently predict the shape of the nonthermal ion (i.e., proton) spectrum up to the neutrino-producing TeV range.
However, regardless of the exact shape of the ion spectrum, our simulations establish that a significant fraction of the coronal
power can be converted into cosmic rays. In particular, we find that ions in the nonthermal tail of the distribution
amount to roughly $q_{\rm i}/4 \sim 1/6$ of the total dissipated power.

In conclusion, we showed that strongly driven and radiative electron-ion turbulence can generate nonthermal ion and electron distributions, while producing
at the same time emission spectra consistent with X-ray observations of black-hole coronae. The steady state is a two-temperature trans-sonic flow
with ions much hotter than electrons. Present computational constraints prevent us from further improving the 
realism of our numerical model (e.g., the box size). A promising future direction may be to disentangle the 
subject into a set of complementary questions, such as the physics of the MeV tail or nonthermal proton acceleration, which can be investigated 
with simplified models specifically designed to address a particular challenge.

\begin{acknowledgments}
We gratefully acknowledge helpful discussions with F.~\mbox{Bacchini}, 
L.~\mbox{Sironi}, R.~\mbox{Mbarek}, B.~\mbox{Ripperda}, H.~\mbox{Krawczynski}, 
A.~\mbox{Hankla}, E.~\mbox{Gorbunov}, D.~\mbox{Caprioli}, L.~\mbox{Comisso}, A.~\mbox{Levinson}, 
M.~\mbox{Nhat Ly}, M.~\mbox{Lemoine}, and C.~\mbox{Nagele}.
D.G.~is supported by the Research Foundation--Flanders (FWO) 
Senior Postdoctoral Fellowship 12B1424N. This work was supported by NASA (grant 80NSSC22K1054) and the Simons Foundation (grant 00001470; A.P.), and was facilitated by the Multimessenger Plasma Physics Center (MPPC; A.P.\ and A.B.) under NSF grant No.~PHY-2206607. A.P.\ additionally acknowledges support from an Alfred P.~Sloan Fellowship and a Packard Foundation Fellowship in Science and Engineering. A.M.B.~is supported by NASA ATP grant 80NSSC24K1229, NSF
grant AST-2408199, and Simons Foundation award 446228. The resources and services used in this work were 
provided by the VSC (Flemish Supercomputer Center), funded by the 
FWO and the Flemish Government.
We acknowledge LUMI--BE for awarding 
this project access to the LUMI supercomputer, owned by the 
EuroHPC Joint Undertaking, hosted by CSC (Finland) and the LUMI 
consortium, through a LUMI--BE Regular Access call.
Simulations were additionally performed on NASA Pleiades (GID s2056).
\end{acknowledgments}
{\software{\textsc{Tristan-MP v2} \citep{tristanv2}, \textsc{xspec} \citep{XSPEC}}}

\appendix

\section{Analysis of Shocks and Current Sheets}
\label{sec:injection}

\begin{figure}[htb!]
\centering
\includegraphics[width=\columnwidth]{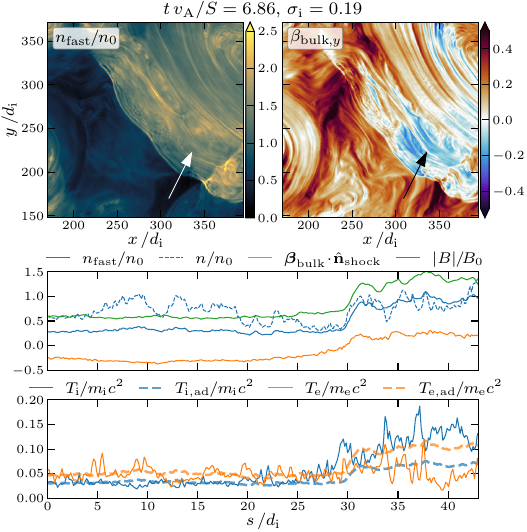}
\caption{\label{fig:shock} Structure of a shock inside the turbulent box of the run with $\sigma_{\rm i}=0.19$. 
Top panels show the fast-mode density and the $y$-component of the ion bulk velocity in a 
fraction of the domain. 1D shock profiles (middle and bottom panel) 
are extracted along the path indicated with arrows. In the middle panel we show the total density $n$, the fast-mode density $n_{\rm fast}$, 
the magnetic field magnitude $B$, and
the bulk velocity component parallel to the shock normal. 
In the bottom we show profiles of the ion
and electron kinetic temperature (solid curves), along with expected values for adiabatic compression (dashed curves).}
\end{figure}

\paragraph{Shock structure}In Fig.~\ref{fig:shock} we analyze the structure of a representative shock extracted from the turbulent volume of our simulation.
In the top panels we visualize the structure of the fast-mode density and the $y$-component of the ion bulk velocity. Along the path indicated by the arrows
we extract 1D shock profiles, which are shown in the bottom two panels of Fig.~\ref{fig:shock}. 
The profiles shown in the middle panel reveal a jump (near $s/d_{\rm i} \approx 30$) in the magnetic field and fluid velocity, which can be associated with the location of the 
shock. The shock transition is a few ion skin depths wide.
The profile of the total density shows the corresponding jump less clearly than the density of 
fast-mode fluctuations alone (see Appendix~\ref{sec:modes}), since the total density includes pressure-balanced fluctuations not belonging to 
fast modes.\footnote{Note that for the pressure-balanced fluctuations 
$\delta n/n_0 \simeq - \alpha^{-1}\,\delta B/B_0$ (see Appendix~\ref{sec:modes}), with $\alpha^{-1} \approx 3.1$ at the 
time analyzed in Fig.~\ref{fig:shock}. Therefore, the pressure-balanced fluctuations show up more clearly in the profiles of $n$ rather than $B$.} 
Based on the fast-mode profile, we determine a 
density compression factor $r\approx 3.2$ between the shock upstream ($s/d_{\rm i} < 30$) and downstream ($s/d_{\rm i} > 30$).
The velocity jump across the shock transition is $\delta v \approx 0.53 c$. The corresponding Alfv\' enic Mach number is
$M_{\rm A} = \delta v/v_{\rm A,u}\approx 2.3$, 
where $v_{\rm A,u}$ is the upstream Alfv\' en speed. The same value is found for the sonic Mach number $M_{\rm s} = \delta v/c_{\rm s,u}$,
where $c_{\rm s,u}$ is the upstream sound speed. The shock analyzed in Fig.~\ref{fig:shock} is quasi-perpendicular; 
the angle of the upstream $\vec B$ field relative to the shock normal is $\approx70^\circ$. In general, the shocks embedded into the turbulent box
feature a broad range of upstream magnetic field orientations, but the quasi-perpendicular case seems to be the most typical.

\begin{figure*}[htb!]
\centering
\includegraphics[width=\textwidth]{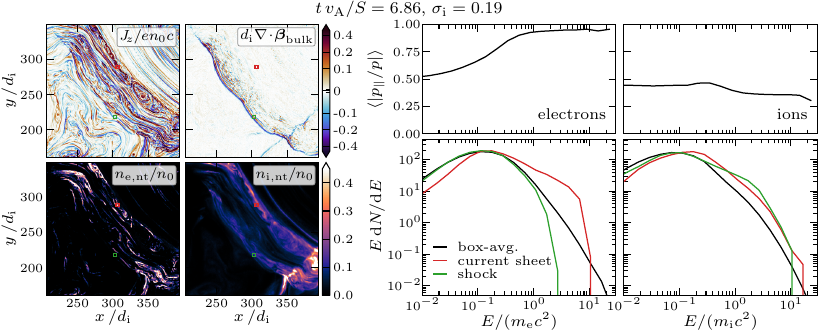}
\caption{\label{fig:injection} Injection of particles into nonthermal 
populations near shocks and current sheets. On the top left we show  the out-of-plane electric current $J_z$ and the 
divergence of the bulk velocity $\nabla\cdot{\boldsymbol\beta}_{\rm bulk}$ in a fraction of the 2D simulation domain. On the bottom left we show the 
density of nonthermal electrons $n_{\rm e,nt}$ and nonthermal ions $n_{\rm i,nt}$. The red and green squares mark the representative 
locations of a current sheet and a shock, respectively, from which we extract the local particle distributions shown on the bottom right side of the figure. On the top right side we 
show the mean cosine of the ion and electron pitch angle relative to $\vec B$, computed in the local $E\times B$ frame.}
\end{figure*}

In the bottom panel of Fig.~\ref{fig:shock} we show the profiles of the ion and electron proper kinetic temperature. For reference,
we also calculate the profiles corresponding to adiabatic 
compression $T_{s,{\rm ad}} = T_{s,{\rm u}}(n_{\rm fast}/n_{\rm fast,u})^{\Gamma_{s}-1}$, where the subscript ``$\rm u$'' denotes the mean upstream
values and $\Gamma_s$ is the species adiabatic index (we use $\Gamma_s = 5/3$ for both species). 
The electrons do not show any appreciable temperature increase beyond expectations for 
adiabatic compression. In fact, the values of $T_{\rm e}$ are even somewhat below $T_{\rm e,ad}$, which may be due to the rapid radiative cooling 
experienced by the electrons. On the other hand, the downstream ion kinetic temperature raises above the expectations for adiabatic compression, which
indicates entropy production across the shock transition.

\paragraph{Particle energization at shocks and current sheets}In Fig.~\ref{fig:injection} we analyze the local particle distributions 
at shocks and current sheets in the simulation with $\sigma_{\rm i}=0.19$. Current sheets can be identified as intense sheetlike 
structures in the out-of-plane current density $J_z$, while shocks can be seen as near-discontinuities in the divergence of the ion 
bulk velocity $\nabla\cdot{\boldsymbol\beta}_{\rm bulk}$ 
(top left side of Fig.~\ref{fig:injection}). Using the spatially resolved particle energy distributions $f_s(E,x,y)$, we analyze how nonthermal features correlate
with shocks and current sheets. On the bottom left side of Fig.~\ref{fig:injection} we show the local structure of the nonthermal electron $n_{\rm e,nt}$ 
and nonthermal ion $n_{\rm i,nt}$ particle density. Nonthermal particles are defined here as those with $E > 4 \langle E_s\rangle$, where 
$\langle E_s\rangle$ is the box-averaged species mean energy
($\approx 0.13\, m_{\rm i}c^2$ for ions and $\approx 0.19\, m_{\rm e}c^2$ for electrons). By comparison with snapshots of $J_z$ and 
of $\nabla\cdot{\boldsymbol\beta}_{\rm bulk}$, we find that nonthermal electron populations are enhanced near regions of intense current, 
whereas the ions display higher nonthermal particle fractions in the vicinity of both shocks and current sheets.

On the bottom right side of Fig.~\ref{fig:injection} we show examples of 
local particle distributions in the vicinity of a shock and a current sheet. The local distributions are normalized such that the integral
$\int f_s(E){\rm d}E$ matches the box-averaged result.
The analyzed locations are indicated with red squares (for the current sheet) and 
green squares (for the shock) in the left panels of Fig.~\ref{fig:injection}. We made sure that the selected point near a shock does not 
accidentally coincide with a current sheet (and vice versa). The local analysis shows that the electron spectrum near a current sheet 
can be quite hard (for the particular case shown $p\approx 1.8$). On the other hand, at the location of the shock electrons show
no obvious signs of enhanced energization. The ion spectrum shows an above-average amount of energetic particles at both the 
location of the shock and the current sheet. Particularly at the shock, the local nonthermal ion spectrum is notably harder than the boxed-averaged result.  
We inspected various locations near shocks and reconnecting current sheets and found local particle 
distributions broadly similar to those shown on the bottom right side of Fig.~\ref{fig:injection}.

Finally, on the top right side 
of Fig.~\ref{fig:injection} we show the energy dependence of the mean cosine of the particle pitch angle $\mu = \langle |p_\parallel/p|\rangle$, 
where $p_\parallel$ is the momentum parallel to the local $\vec B$ field and $p$ is the total momentum. 
Following previous works \citep{Comisso2019}, we compute $\mu$
in the local $E\times B$ frame. In the nonthermal parts of the distribution, the electrons closely align their 
momenta with the field-parallel direction. This provides additional evidence that electrons are primarily energized
at reconnecting current sheets, and in particular via the nonideal $E_\parallel$ fields \citep[see also][]{Comisso2019,Comisso2024}. 
In contrast, the ion momentum distribution is more isotropic and shifts slightly toward lower values of $\mu$ with growing energy.

\begin{figure}[htb!]
\centering
\includegraphics[width=\columnwidth]{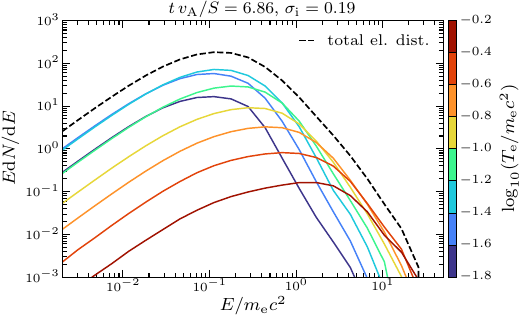}
\caption{\label{fig:el_spec_te} Electron energy distributions extracted from locations with different values of the local temperature. 
Each color curve represents the distribution of particles from spatial 
cells where $\log_{10}(T_{\rm e})$ falls into a given range (as indicated by the colorbar). The distributions are normalized relative to the number of particles 
contained in a given band, such that their sum gives the total (box-averaged) distribution (black dashed curve).}
\end{figure}

\paragraph{Composite electron spectrum}In Fig.~\ref{fig:el_spec_te} we show electron distributions extracted from spatial cells
where the local value of $T_{\rm e}$ falls into a given range.
This further clarifies how the total spectrum is shaped by turbulent fluid motions of moderately hot electrons in the bulk of the volume, 
and by nonthermal particles at spatially localized hotspots.

As shown in Fig.~\ref{fig:el_spec_te}, the most numerous particle population (cyan curve) has a typical proper kinetic temperature of 
$T_{\rm e} \approx 10^{-1.3}m_{\rm e} c^2 \approx 0.05\,m_{\rm e} c^2$. Such temperature is too low to fully account for the peak of the box-averaged 
spectrum near $E \approx 0.15 m_{\rm e}\,c^2$. The discrepancy can be explained by contributions from bulk turbulent motions. Indeed, even in the lowest temperature band (dark blue curve) 
with $T_{\rm e} \approx 10^{-1.7}m_{\rm e} c^2 \approx 0.02\,m_{\rm e} c^2$ the peak of the distribution is near $E \approx 0.1m_{\rm e} c^2$, which indicates that bulk motions put an effective
floor on the minimum particle energy. In other words, in the simulation frame the electron energies are boosted by the bulk motions, and therefore
the distributions peak at larger values than expected for a given temperature.

At the most extreme hotspots with $T_{\rm e} \approx 10^{-0.3}m_{\rm e} c^2 \approx 0.5\,m_{\rm e} c^2$ (dark red curve in Fig.~\ref{fig:el_spec_te}), the electron distribution exhibits a 
hard power law for $0.1\,m_{\rm e} c^2 \lesssim E \lesssim  2m_{\rm e}c^2$, and a steeper spectrum above this range.
The hard power-law range (with $p\lesssim 1$) is reminiscent of spectra obtained in
high-$\sigma_{\rm e}$ reconnection studies \citep{Guo2024,Sironi2025}. Together with the results shown in Fig.~\ref{fig:injection}, we can conclude that turbulent current sheets
can produce spectra broadly similar to those expected from dedicated reconnection studies \citep[e.g.,][]{Comisso2024}. 
However, because the most intense energization sites occupy a small volume, the composite (box-averaged)
electron spectrum is significantly softer and different from the spectrum of an isolated current sheet.

\section{Photon-photon and Ion-electron Collision Rates}
\label{sec:collisions}

Apart from Compton scattering, our simulations do not account for other binary particle interactions, 
such as Coulomb collisions or two-photon pair production, 
which may be relevant in black-hole coronae \citep{Fabian2015, Bambic2024a, Nattila2024}. Here we estimate the expected rates of these interactions.

\paragraph{Pair production}With growing compactness $\ell$, the system can transition into a regime with
copious electron-positron pair production, which would invalidate our assumed electron-ion 
composition. The mean steady-state positron density $n_+$ can be estimated by balancing pair production and annihilation \citep{Svensson1987,Beloborodov2017}:
\begin{align}
    \dot{n}_{\gamma\gamma} \simeq \eta\sigma_{\rm T}cn_1^2 \simeq \dot{n}_{\rm ann}\simeq (3/8)\sigma_{\rm T}c n_+(n_0 + n_+),
\end{align}
where $\eta \simeq 0.1$ and $n_1 = f_1 n_{\rm ph}$ is the density of photons with energies $> m_{\rm e}c^2$.
By introducing the pair-loading factor $Z_\pm = 2 n_+ / n_0$, we can estimate 
for $\eta\simeq 0.1$ that
\begin{align}
Z_\pm \simeq \left[1 + (f_1 n_{\rm ph}/n_0)^2\right]^{1/2} - 1.
\end{align}
Using \eqref{eq:nph} we find that $Z_\pm \gtrsim 1$ when:
\begin{align}
   \sigma_{\rm i} \gtrsim \left[\frac{\sqrt{12}E_0(A - 1)}{f_1 m_{\rm i}c^2(1-q_{\rm i})(\tau_{\rm T}+1)} \right]^{2/3}.
    \label{eq:pp}
\end{align}
The measured fraction of photons with $E> m_{\rm e}c^2$ 
in our runs with $\ell \approx 0.6, 2.5, 8.5$ 
is $f_1\approx 3.4\!\times\! 10^{-6}, 6.5\!\times\! 10^{-6}, 7.4\!\times\! 10^{-6}$, respectively. 
With a conservative choice of $f_1 = 10^{-5}$, we find for a representative 
set of parameters ($m_{\rm i}/m_{\rm e} = 1836$,
$q_{\rm i} = 0.65$, $\tau_{\rm T} = 1$, $E_0 = 2.7\times10^{-5}m_{\rm e}c^2$, 
and $A-1 = 16$) that condition \eqref{eq:pp} amounts to $\sigma_{\rm i} \gtrsim 0.24$. According to 
\eqref{eq:sigma_range}, significant pair production is therefore expected only at the high 
end of the plausible $\sigma_{\rm i}$ range, where $\ell \sim 100$. Note that the mapping between
$\sigma_{\rm i}$ and $\ell$ is artificially modified in our simulations, owing to the reduced 
ion-electron mass ratio. For this reason, our largest simulated $\ell\approx 8.5$ is far below the range
for pair production, even though the corresponding $\sigma_{\rm i}=0.19$ might be relatively 
close to the $Z_\pm \sim 1$ boundary for a realistic $m_{\rm i}/m_{\rm e}$. 

\paragraph{Ion-electron collisions}Rapid Coulomb collisions between ions and electrons can lead to 
equilibration of their temperatures. It is therefore useful to check when collisions are sufficiently
infrequent to allow for the development of a two-temperature corona with $T_{\rm i}\gg T_{\rm e}$.
For simplicity, we employ analytical estimates for a Maxwellian plasma 
with dimensionless species temperatures $\theta_{\rm s} = T_{\rm s}/(m_{\rm s}c^2)$ \citep{Stepney1983, Stepney1983b}, where in practice $\theta_{\rm s}$ is to be understood as an effective 
kinetic temperature of the (nonthermal) particle distribution.
The collisional ion-electron relaxation time scale, on which hot ions lose energy to 
electrons,  can be obtained 
as $t_{\rm ie} \simeq \left| \kappa_{\rm e} n_0(T_{\rm i} - T_{\rm e}) / ({\rm d}U_{\rm e}/{\rm d} t)\right|$
\citep{Spitzer1962}, where ${\rm d}U_{\rm e}/{\rm d} t$ is the collisional electron energy 
exchange rate per unit volume, and $\kappa_{\rm e} = 1 / (\Gamma_{\rm e} - 1)$. 
For non-relativistic or very mildly relativistic particles we can approximate \citep{Stepney1983}:
\begin{align}
    \frac{t_{\rm ie}}{s_{\rm esc} / v_{\rm A}} \simeq \frac{\sqrt{\pi}(m_{\rm i}/m_{\rm e})(v_{\rm A}/c)}
    {\sqrt{2}\tau_{\rm T}\log\Lambda}\left(\theta_{\rm e} + \theta_{\rm i}\right)^{3/2},
\end{align}
where the Coulomb logarithm $\log\Lambda\simeq 20$. Using 
$\theta_{\rm i} \simeq \sigma_{\rm i}/\Gamma_i$ (per Eq.~\eqref{eq:Ti}), we find for $\theta_{\rm e} = 0.1$ and 
$m_{\rm i} / m_{\rm e} = 1836$ that the ion-electron collisional equilibration time is longer than the dynamic time 
($t_{\rm ie} > s_{\rm esc}/v_{\rm A}$) when $\sigma_{\rm i} \gtrsim 0.04$.

\section{Turbulent Mode Decomposition}
\label{sec:modes}

Further insight into the nature of the turbulent cascade can be obtained by decomposing the
turbulent-energy fluctuations among different linear modes supported by the plasma. 
The identification of modes is based on their polarization properties, which are 
reasonably well preserved even in strongly turbulent regimes \citep{Groselj2019}.
Here, we consider the partitioning of energy between 
the MHD Alfv\' en, fast, and slow mode. Apart from its simplistic appeal, the use of MHD relations 
is supported by studies showing that the large-scale dynamics of turbulent kinetic plasmas is often
remarkably MHD-like \citep[e.g.,][]{Verscharen2017,Meyrand2019,Squire2019}.

Our 2D simulation domain is limited to wave vectors $\vec k_\perp$ perpendicular to the
out-of-plane mean magnetic field $\vec B_0 = B_0\hat{\vec z}$. However, an \emph{effective} field-parallel wave number $k_\parallel$ is 
still retained via the projection of $\vec k_\perp$ onto the \emph{local} magnetic 
field \citep[e.g.,][]{Li2016,Groselj2017}. Under these circumstances, it is most reasonable to consider 
linear modes in the limit of quasi-perpendicular 
propagation \citep[see][for details]{Schekochihin_notes}. In this limit, the decomposition becomes 
independent of $k_\parallel$, as shown below. For full 3D mode decomposition at 
arbitrary $\vec k$, see e.g.~\citet{Cho2003}.

To lowest order in $k_\parallel/k_\perp\ll 1$, the polarizations are determined as follows.
For a given $\vec k \approx \vec k_\perp$, shear Alfv\' en modes 
feature magnetic field and velocity fluctuations aligned with 
the $\vec B_0 \times \vec k_\perp$ direction, while the electric field is  
aligned with $\vec k_\perp$. The fast mode has velocity vector aligned with
$\vec k_\perp$, and its electric field is parallel to $\vec B_0 \times \vec k_\perp$.
The velocity fluctuation of the slow mode is along $\vec B_0$. 
Fluctuations of the plasma density $\delta n$ and magnetic 
field magnitude $\delta B\approx \delta B_z$ are contained in both the slow and fast mode. Introducing the 
variables $\delta \tilde p = 4\pi\delta p/B_0$ and $\delta\tilde n = B_0\delta n/n_0$, 
where $\delta p$ is the pressure fluctuation, we can decompose $\delta B$ as follows:
\begin{align}
    \delta B & = \delta B_{\rm slow} + \delta B_{\rm fast}, & \\
    \delta B_{\rm slow} & = (1 + 1/\alpha)^{-1}(\delta B - \delta\tilde n) = -\delta\tilde p_{\rm slow},
   \label{eq:slow} 
    & \\
    \delta B_{\rm fast} & = (1 + \alpha)^{-1}(\delta B + \alpha\delta\tilde n) = \delta\tilde n_{\rm fast},
    \label{eq:fast}&
\end{align}
where the pressure fluctuation is related via the adiabatic response to the density as
$\delta\tilde p = \alpha \delta\tilde n$, $\alpha = c_{\rm s}^2 / v_{\rm A}^2$ 
is the squared ratio of the sound speed $c_{\rm s}$ to the Alfv\' en velocity $v_{\rm A}$, the ``slow'' subscript
denotes the slow mode contribution, and the ``fast'' subscript indicates 
the part identified as the fast mode. Eq.~\eqref{eq:slow} shows that the slow mode fluctuations are 
essentially pressure-balanced,  whereas Eq.~\eqref{eq:fast} implies conservation of magnetic flux by the fast-mode 
plasma compressions. The turbulent energy at a given $\vec k_\perp$ can be then written as a sum of the 
3 modes:
\begin{align}
\mathcal{E}(\vec k_\perp) = \, & \mathcal{E}_{\rm Alfven}(\vec k_\perp) + \mathcal{E}_{\rm slow}(\vec k_\perp) +
\mathcal{E}_{\rm fast}(\vec k_\perp) \nonumber\\
=\, & \frac{1}{8\pi}\left\{|\vec B(\vec k_\perp)|^2 + |\vec E_\perp(\vec k_\perp)|^2\right\} \nonumber \\
  & + |{\boldsymbol \upsilon}(\vec k_\perp)|^2 + \frac{\alpha}{8\pi}|\delta\tilde n(\vec k_\perp)|^2, \label{eq:turb_ene}
\end{align}  
\begin{align}
    \mathcal{E}_{\rm Alfven}(\vec k_\perp)  = \, &\frac{1}{8\pi}|\vec B_\perp(\vec k_\perp)|^2 +  |(\hat{\vec z}\times\hat{\vec k}_\perp)\cdot {\boldsymbol \upsilon}_\perp(\vec k_\perp)|^2\nonumber\\
    & + \frac{1}{8\pi}|\hat{\vec k}_\perp\cdot\vec E_\perp(\vec k_\perp)|^2,
\end{align}
\begin{align}
 \mathcal{E}_{\rm slow}(\vec k_\perp) =\,  &  \frac{1 + 1/\alpha}{8\pi} |\delta B_{\rm slow}(\vec k_\perp)|^2 + |\upsilon_z(\vec k_\perp)|^2,
\end{align}
\begin{align}
 \mathcal{E}_{\rm fast}(\vec k_\perp) =\, & \frac{1+\alpha}{8\pi}|\delta B_{\rm fast}(\vec k_\perp)|^2 
 + |\hat{\vec k}_\perp\cdot {\boldsymbol \upsilon}_\perp(\vec k_\perp)|^2 \nonumber\\
  & + |(\hat{\vec z}\times\hat{\vec k}_\perp)\cdot \vec E_\perp(\vec k_\perp)|^2, 
\end{align}
where $\hat{\vec k}_\perp = \vec k_\perp / k_\perp$,
$\boldsymbol{\upsilon} = [\Gamma_{\rm bulk}^3w_0/(\Gamma_{\rm bulk} + 1)]^{1/2}{\boldsymbol\beta}_{\rm bulk}$ is the weighted fluid velocity, 
$\Gamma_{\rm bulk} = [1 - \beta_{\rm bulk}^2]^{-1/2}$ is the fluid Lorentz factor, and $w_0$ is the mean 
proper enthalpy density.\footnote{For completeness, we retain the electric field contribution and we use 
a relativistic definition of the bulk kinetic energy density, but we note that relativistic effects are mild in the
regime of interest.} Note that $|{\boldsymbol \upsilon}|^2 = \Gamma_{\rm bulk} w_0 (\Gamma_{\rm bulk} - 1)$, which 
reduces to $n_0 m_{\rm i}c^2 \beta_{\rm bulk}^2/2$ in the non-relativistic limit.
The last term in Eq.~\eqref{eq:turb_ene} can be written in
real space as $\delta n\delta p/(2n_0)$. It can be identified 
as the energy density of acoustic oscillations, which naturally enter the total energy definition
when the cascade is compressible \citep[e.g., see][]{Schekochihin2009}.

The above-described MHD mode decomposition is applied to our simulation data in spectral space and in real space. 
Since the ions are much hotter than electrons, the ion pressure and density 
are used to obtain $\delta\tilde p$ and $\delta\tilde n$, respectively. The ratio $\alpha$ is estimated at each given time from the root-mean-square
fluctuations of the pressure and density as $\alpha = \delta\tilde p_{\rm rms} / \delta\tilde n_{\rm rms}$.
Given the large-amplitude nature of our turbulence drive, the mode decomposition is not exact, even within the scope of MHD. 
In particular, the direction parallel to the local magnetic field $\hat{\vec b} = \vec B/B$  can significantly deviate from $\vec B_0$. 
In real space, the fluctuating fields can be readily projected onto the local field-parallel direction. Therefore, 
in real space we use $\delta B \approx \delta(\hat{\vec b}\cdot\vec B)  = \delta|B|$ when extracting the fast mode fluctuations.
On the other hand, in spectral space the local projection introduces a certain ambiguity, because it convolves
modes with different field components and wavenumbers. We therefore prefer to use $\vec B_0$ as the field-parallel direction 
for the spectral decomposition. Nevertheless, we confirmed that similar results are obtained when the 
parallel direction is defined with respect to $\hat{\vec b}$.

\begin{figure}[htb!]
\centering
\includegraphics[width=\columnwidth]{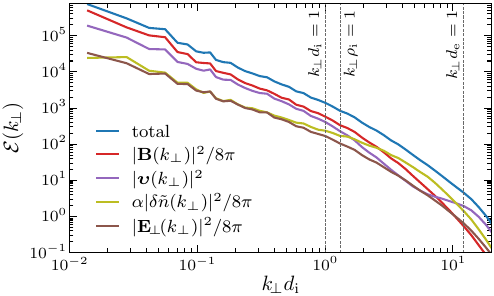}
\caption{\label{fig:spec_appendix} Total turbulent energy spectrum from Fig.~\ref{fig:turb_spectra} decomposed
among the different forms of energy that appear in Eq.~\eqref{eq:turb_ene}.}
\end{figure}

For reference, we demonstrate in Fig.~\ref{fig:spec_appendix} how the 
total energy spectrum (as shown in Fig.~\ref{fig:turb_spectra} of Sec.~\ref{sec:turbulence}) 
is partitioned among the different forms of energy that appear in Eq.~\eqref{eq:turb_ene}. 
As expected, at MHD scales ($k_\perp \lesssim 1/d_{\rm i}$) most of the energy is contained in the magnetic (red curve) and bulk kinetic energy 
density (purple curve). At sub-ion scales, the energy is primarily contained in the magnetic and density fluctuations (yellow curve), consistent with
expectations for kinetic Alfv\' en turbulence \citep[e.g.,][]{Groselj2019}. The flattening of the kinetic energy density 
spectrum at $k_{\perp}d_{\rm i} \sim 10$ can be attributed to PIC noise artifacts.

\bibliographystyle{aasjournal}

\end{document}